\documentclass[aps,prb,reprint, superscriptaddress, twocolumn, longbibliography]{revtex4-2}
\usepackage{float}
\usepackage{graphicx} 
\usepackage{dcolumn}  
\usepackage{bm}       
\usepackage{verbatim}
\usepackage{physics}  
\usepackage{multirow}
\usepackage{xcolor}
\usepackage{hyperref} 
\usepackage{amssymb} 
\usepackage{booktabs}
\usepackage{siunitx}
\newcommand\ddfrac[2]{\frac{\displaystyle #1}{\displaystyle #2}}

\definecolor{edit}{HTML}{C0445A}

\renewcommand{\eqref}[1]{Eq.~(\ref{#1})}

\usepackage[normalem]{ulem}

\newcommand{\lsi}{LSI, CNRS, CEA/DRF/IRAMIS, \'Ecole Polytechnique, Institut Polytechnique de Paris, F-91120 Palaiseau, France}
\newcommand{\etsf}{European Theoretical Spectroscopy Facility (ETSF)}
\newcommand{\soleil}{Synchrotron SOLEIL, L'Orme des Merisiers, Saint-Aubin, BP 48, F-91192 Gif-sur-Yvette, France}
\newcommand{\bochum}{Research Center Future Energy Materials and Systems of the University Alliance Ruhr and Interdisciplinary Centre for Advanced Materials Simulation, Ruhr University Bochum, Universitätsstraße 150, D-44801 Bochum, Germany}

\begin{document}

\title{Designing explicit functionals for the charge density in terms of a potential}

\author{Muhammed H. Güneş}
\affiliation{\lsi}%
\affiliation{\etsf}%

\author{Ayoub Aouina}
\affiliation{\bochum}
\affiliation{\etsf}%

\author{Vitaly Gorelov}
\affiliation{\lsi}%
\affiliation{\etsf}%

\author{Matteo Gatti}
\affiliation{\lsi}%
\affiliation{\etsf}%
\affiliation{\soleil}

\author{Lucia Reining}
\affiliation{\lsi}%
\affiliation{\etsf}%
\email{lucia.reining@polytechnique.fr}

\date{\today}

\begin{abstract}
One of the most powerful strategies to address properties of real many-body systems is to incorporate data obtained for models, for example, to use data of the homogeneous electron gas in order to build the Local Density Approximation for the Kohn-Sham exchange-correlation potential. In the present work, we examine to what extent we can use model data to design functionals directly for observables of materials. In particular, we study different approximations for the charge density of real inhomogeneous materials expressed as a simple, explicit functional of a given Kohn-Sham potential, using as central building block the Lindhard density-density response function of the homogeneous electron gas. Our increasingly realistic set of approximations includes a fully nearsighted expression equivalent to the Thomas-Fermi approximation, functional Taylor expansions, and different approximations to the Connector Theory developed in [Aouina \textit{et al.}, npj Computational Materials {\bf 11}, 242 (2025)]. In all cases, the charge density is obtained without ever solving the Kohn-Sham Schr\"odinger equation. Results for cubic helium, a prototypical strongly inhomogeneous material, as well as the covalent semiconductor silicon and metallic aluminum, systematically improve with higher levels of approximation. At the present stage, the results may be used for qualitative discussions or as optimized starting point for a self-consistent Kohn-Sham cycle. More generally, their quality indicates that this is a promising route to obtain functional expressions for observables that are relatively simple to calculate and to analyze.

\end{abstract}

\maketitle

\section{Introduction}

Many properties of materials can be formulated as expectation values of the respective operators using the many-body ground-state wave function. However, to obtain and to store this wave function can be very difficult for large systems, and prohibitive when interaction effects are taken into account \cite{lecture}. One way to overcome the problem is to express quantities of interest as functionals of objects that are more compact than the many-body wave functions. Usually, these functionals are unknown and have to be approximated. One of the most well-known examples is Density Functional Theory (DFT) \cite{Hohenberg1964}, where much effort goes into expressing the total energy approximately as a functional of the ground-state charge density. Since the charge density itself is a ground state observable that has to be calculated, Kohn and Sham introduced an auxiliary system that yields the charge density in principle exactly, while in practice the auxiliary potential of this system is unknown \cite{Kohn1965}. It is in turn a functional of the charge density, which makes the problem self-consistent. Similarly, observables can be expressed as functionals of the one-body Green's function that is obtained from a Dyson equation, where the self-energy plays the role of the auxiliary potential. In turn, this self-energy is usually approximated as a functional of the one-body Green's function itself \cite{Martin2016}. This strategy is very successful, but it is also very intricate \cite{Teale2022}, which means that many flavors of approximations exist that are hard to compare. It also implies that it is often difficult to interpret results, and it also means that inverting the problem to find materials with desired properties is essentially out of reach \cite{Zunger2019}. 

In the present work, we will take a different, more direct direction: the long-term aim is to express observables directly as explicit functionals of the external potential, without intermediate objects or auxiliary systems. This route to build ``potential functionals'' has been evoked before as an alternative to DFT, and in this context strategies for approximations were proposed \cite{Kohn1965,Englert1984,Englert1992,Yang2004,Gross2009,Elliott2008,Cangi2010,Cangi2011,CANGI2013,Witt2019}, but it turns out that finding reasonably simple approximations to the energy for real three-dimensional materials that would perform better than the simple Thomas-Fermi expression remains a challenge. Such potential functionals avoid the calculation of orbitals, just as in orbital-free DFT, which started from the work of Thomas and Fermi \cite{Thomas1927,Fermi1928}. More recently, they motivate strategies to machine learn the energy and charge density. An overview can be found in \cite{Mi2023}. It should be stressed that in all cases the aim is not to obtain the exact expressions, since these are known, but exceedingly difficult to evaluate in practice \footnote{Any observable $O$ in equilibrium can be expressed as a trace involving the operator $\hat O$ of the observable and the many-body Hamiltonian, 
$$
    O=\frac{1}{Z}\mathrm{Tr}\!\left( e^{-\beta \hat H}\,\hat O \right),
$$
with \(Z = \mathrm{Tr}\; e^{-\beta \hat H}\) the partition function and \(\beta = 1/(k_B T)\). }: the question is how to obtain the most useful compromise between accuracy and computational cost. 

To find manageable expressions, we can look back to what has turned out to be successful before. In particular, both in the context of density functionals and of Green's function functionals, a crucial breakthrough was obtained by incorporating results from solvable model systems. The Local Density Approximation (LDA) \cite{Kohn1965} of DFT is based on Quantum Monte Carlo results for the homogeneous electron gas \cite{Ceperley1980} (HEG), and the single-site approximation to Dynamical Mean Field Theory uses the Anderson Impurity Model results to build a local self-energy \cite{Georges1996}. The case of the LDA is particularly appealing, since the HEG results were produced once and for all, and they have been used innumerable times since then. The idea was generalized recently by the Connector Theory (COT) approach, where a quantity of interest is expressed as a functional that involves the result of a model with suitably chosen parameters, and proof-of-principle illustrations for simple systems were given \cite{Vanzini2022}. 
COT was then successfully applied to design new density functional approximations for the exchange-correlation (xc) potential, the unknown part of the auxiliary Kohn-Sham (KS) potential \cite{Aouina2025}. 
The resulting approximate Kohn-Sham potential was used to determine the charge density from the solutions of the Kohn-Sham equation. However, the approach is not limited to the construction of density functionals.

In the present work, our goal is to approximate \textit{directly} the charge density, without \textit{solving} the Kohn-Sham equations. We will do this by expressing the charge density as a simple functional of the total Kohn-Sham potential. The specific question that we address is to what extent the use of model data, here from the homogeneous electron gas, may help the construction of powerful potential functionals. The simplest way to use knowledge from the HEG is to expand the real system around a homogeneous one. This idea was explored early on. It requires knowledge of the static density-density response functions of the HEG. The first order response is given by the Lindhard function \cite{Lindhard1954}, higher orders were provided later by various authors for the non-interacting gas \cite{Lloyd1968,Brovman1971,Brovman1972,Hammerberg1974,Pickenhain1976,Milchev1977,Porter2010,Witt2019} and beyond \cite{Paasch1977,Paasch1980,Heinrich1980}. The validity of a Taylor expansion of the density was tested mainly for soft, empirical pseudopotentials, and it was found that a straightforward first order expansion is in general unsatisfactory \cite{Vinsome1971}. It has been shown that adding the second order correction leads to significant improvement for mild potentials and slowly varying densities \cite{Heinrich1980}, whereas with increasing potential strength, ionicity, or covalency, results degrade \cite{Baldereschi1981,Heinrich1980}, and it is difficult to improve upon the simple Thomas-Fermi approximation \cite{Pickenhain1976,Milchev1977}. This points to a strong need for at least approximate summation to infinite order if such approaches are to be used for materials with potentials of bare nuclei or with the pseudopotentials used in current \textit{ab initio} calculations, and in general, for materials with strongly inhomogeneous densities. We explore two ways to achieve this, using an approximate termination of the Taylor series due to Schl\"omilch, Cauchy and Lagrange on one side, and COT in combination with Taylor expansions on the other side, and we combine the two ways into a powerful scheme. 

Our aim here is to avoid the solution of the Kohn-Sham equation by using the response to the total Kohn-Sham potential, which involves the response functions of a non-interacting system.
One could also try to express the charge density as a functional of the \textit{external} potential: in that case, even the Kohn-Sham formalism itself would be avoided. This is of course more ambitious, and it will be left for future work. Here, we explore a first step in this direction, and we take the Kohn-Sham potential as given, which means that we work with a chosen approximate density functional and that we do not explicitly include interaction effects in our potential functional.

In the next section, Sec. \ref{sec:framework}, we will briefly introduce the Taylor expansion of the density expressed as a functional of the KS potential. We will then explain how COT applies to our problem, how a Taylor expansion can be used in this framework, and in which way COT will improve the result of the Taylor expansion. To conclude that section, we will also discuss how one can further optimize the resulting approach. In Sec. \ref{sec:results} results will be presented, first as a general illustration in Subsec. \ref{subsec:results-general}, followed by a series of results with increasing accuracy in Subsec. \ref{subsec:results-lpa-start}. Remaining errors are quantified in Subsec. \ref{subsec:quant} for the density itself, and for density-derived observables in Subsec. \ref{subsec:observables}. Calculations are performed for a prototypical strongly inhomogeneous solid, cubic helium, at equilibrium pressure as well as for compressed and for expanded structures, and for silicon and aluminum. Since our model is the HEG, these scenarios constitute a very robust test case, spanning the range from moderately to strongly inhomogeneous densities. In Sec. \ref{sec:ks} we explore how our resulting functionals behave in a loop where the density in the KS potential is updated self-consistently. Conclusions are given in Sec. \ref{sec:conclusions}.  

\section{The framework}
\label{sec:framework}

At the heart of our strategy for the construction of explicit functionals is the systematic re-use of model data that are created once and for all. One simple way to reconstruct results for a given system from a set of model results is to Taylor expand the real system around the model. Taylor expansions will therefore be one of the ingredients in our approach. The second main ingredient is COT, which was introduced in \cite{Vanzini2022} and applied to the design of approximate xc potentials in \cite{Aouina2025}. It uses knowledge from the model system to replace the missing information of a given approximation; here, since we use a low order Taylor expansion, COT implicitly replaces in an optimized way the higher orders in the real system by those of the model. For both the Taylor expansion and COT, we choose as model the HEG. It is described by just one parameter, namely the potential difference from the chemical potential, which determines its density. This choice, as well as the strategy used here to develop the functional $n({\bf r};[v_{\rm KS}])$ for the charge density $n$ in terms of the KS potential $v_{\rm KS}$, are analogous to the strategy used in \cite{Aouina2025} to develop the functional $v_{\rm xc}({\bf r};[n])$ for the KS xc potential $v_{\rm xc}$ in terms of the density $n$, and it is interesting to see that many findings turn out to be general, although the problem is completely different.

\subsection{First-order functional Taylor expansion}

We start with the simplest approximation, a first order Taylor expansion of $n({\bf r};[v_{\bf r}])$ around a potential that is constant, $v_{\bf r}(\tilde {\bf r})=v_{0{\bf r}}$. A \textit{different homogeneous} starting potential $v_{0{\bf r}}$ can be chosen for every point ${\bf r}$ in which we wish to calculate the density; this is expressed by the subscript ${\bf r}$. Note that throughout the manuscript, the use of arguments and subscripts will be such that $f_{\bf r}(\tilde {\bf r})$ is a function $f$ of $\tilde {\bf r}$ that is chosen in some optimal way for a given point ${\bf r}$; if we are interested in a different point ${\bf r}'$, a different function $f_{{\bf r}'}(\tilde {\bf r})$ may be chosen. 

Here and in the following, we will set the chemical potential to zero, since in an extended system only potential differences are important. In principle, there is also the freedom to choose how to align the chemical potentials of the HEG and the real system. We choose this alignment such that densities are always real and positive, while optimizing the number of electrons with respect to the correct, known value (for different choices of chemical potential for cubic helium and silicon, respectively, see Appendix~\ref{app:mu}). In the following, for the sake of compactness, we will denote $v_{\rm KS}\to v$. To first order, the Taylor expansion of the functional reads
\begin{equation}
    n(\textbf{r};[v]) \approx n(\textbf{r};[v_{0{\bf r}}]) + \int d\textbf{r}' \chi_0^h(|\textbf{r}-\textbf{r}'|,v_{0{\bf r}}) \Big( v(\textbf{r}') - v_{0{\bf r}}\Big)\,.
    \label{eq:taylor-1}
\end{equation}
The zero-order term is the Thomas-Fermi expression \cite{Ashcroft1976}
\begin{equation}
   n(\textbf{r};[v_{0{\bf r}}])= n^h(v_{0{\bf r}})\equiv \frac{(-2v_{0{\bf r}})^{3/2}}{3\pi^2}\,,
    \label{eq:TF}
\end{equation}
with $k_F = (-2v_{0{\bf r}})^{1/2}$ parametrically dependent on $\textbf{r}$. The first order term is evaluated with the Lindhard function $\chi_0^h$ \cite{Lindhard1954}, for a density that is determined by $v_{0{\bf r}}$. The Lindhard function can be found in textbooks \cite{Giuliani2005}, so all ingredients are easily accessible.

\subsection{Simulating higher orders: Connector Theory}

\begin{figure}[h!]
    \centering
    \includegraphics[width=\columnwidth]{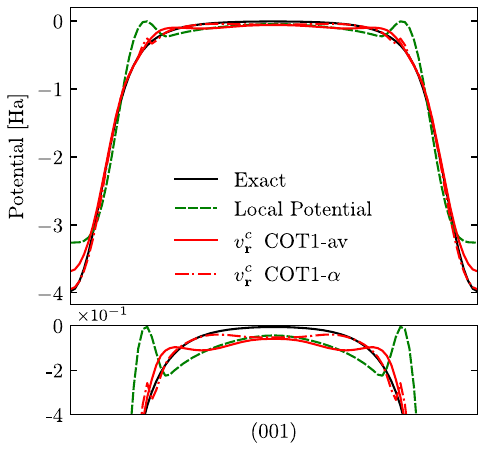}
    \caption{Local Kohn-Sham potential along the (001) direction of cubic helium in the LDA, compared to the connector potential $v^c_{\bf r}$ that would yield exactly the same density, and to the COT1-av and COT1-$\alpha$ approximations for the connector (see Table \ref{tab:all}). Atoms are situated at the left and right sides of the displayed region. The lower panel is a zoom on the low density region.}
    \label{connectors}
\end{figure}

As we will see later, $n^h(v_{0{\bf r}})$ gives a good idea of the density, but it is far from the accuracy that is required in most applications. However, the function $n^h(v)$ spans all possible values of a density. Therefore, there exists a homogeneous potential $v^c_{\bf r}$ for which $n({\bf r})=n^h(v_{\textbf{r}}^c)$, i.e., where the exact density in a given point ${\bf r}$ equals the density of a HEG with potential $v_{\textbf{r}}^{c}$. Fig.~\ref{connectors} illustrates this concept. For cubic helium, the figure shows the LDA Kohn-Sham potential $v({\bf r})$ that enters the Kohn-Sham Schr\"odinger equation to calculate the density $n({\bf r})$ (for computational details, see Sections \ref{sec:results} and Appendix \ref{sec:compdet}). It also shows the exact $v^c_{\bf r}=(n^h)^{-1}(n(\textbf{r}))$, i.e., in every point ${\bf r}$, the magnitude of a homogeneous potential that would produce a homogeneous density with value $n({\bf r})$: this is the exact COT potential. It is overall similar to the Kohn-Sham potential \footnote{Note that here we are working with a local pseudopotential, so the potential itself is not a measurable quantity, it only defines our Hamiltonian. For example, the strong wiggle at the bottom of the atoms appears in the pseudopotential used here, but other choices for the helium pseudopotential are possible that do not show these wiggles. The density, instead, should not depend on the pseudopotential that is used for all reasonable choices, except close to the nucleus. }. However, it is much deeper on the atoms that are situated at the left and right ends of the figure, it rises more steeply, and it then becomes much smoother than the KS potential between the atoms. 

As we will see later, these differences have a significant impact on the density. 

Of course, in practice, the density is unknown and the exact $v^c_{\bf r}=(n^h)^{-1}(n(\textbf{r}))$ cannot be calculated.
The COT approach proposes a strategy to find a good approximation for this \textit{connector} potential $v_{\textbf{r}}^{c}$. Starting from the exact definition, we make a consistent approximation, i.e.,
\begin{equation}
    n({\bf r})=n^{h}(v^{c}_{{\bf r}})\,\,\,\,\,\to\,\,\,\,\,n^{\rm app}({\bf r})\approx n^{h,{\rm app}}(v^{c,{\rm app}}_{{\bf r}})\,.
    \label{eq:cot-strategy}
\end{equation}
The approximation is chosen such that $n^{\rm app}({\bf r})$ can be determined, which allows us to solve the approximate equation for $v^{c,{\rm app}}_{{\bf r}}$ \footnote{The fact that the exact connector exists does not guarantee that the approximate equations also have a solution. This problem is treated on a case-by-case basis.}. This approximation to $v^{c}_{{\bf r}}$, which we denote for clarity as $v^{c,{\rm app}}_{{\bf r}}$ in the present paragraph, is then used in $n^{\rm COT}({\bf r})=n^h(v^{c,{\rm app}}_{{\bf r}})$ to obtain the final COT approximation for the density. Note that it is very important to distinguish where approximations are made and where exact expressions are used. The fact that in the last step the exact expression for $n^h$ appears, instead of $n^{h,{\rm app}}$, means that $n^{\rm COT}({\bf r})\neq n^{\rm app}({\bf r})$, i.e., the result is different from the initial approximation used in \eqref{eq:cot-strategy}. In other words, 
COT adds to the initial approximation corrections taken from a HEG that is optimized locally through the use of $v^{c,{\rm app}}_{{\bf r}}$.

The most straightforward choice for an approximation is the first-order Taylor expansion. With this choice $n^{\rm app}({\bf r})$ is given by \eqref{eq:taylor-1}. For the right hand side of \eqref{eq:cot-strategy} we also need the first-order expansion of $n^h$ around $v_{0{\bf r}}$. It reads 
\begin{equation}
    n^h(v^c_{\bf r}) \approx n^h(v_{0{\bf r}}) +\frac{d n^h(v^h)}{dv^h}\bigg\rvert_{v_{0{\bf r}}} (v^c_{\bf r}-v_{0{\bf r}})\,.
\end{equation}
Note that from here on, we drop the superscript ``$\rm app$" in $v^{c,{\rm app}}_{{\bf r}}$. With
\[
\frac{d n^h(v^h)}{dv^h}\bigg\rvert_{v_{0{\bf r}}} = \int d\textbf{r}' \frac{\delta n(\textbf{r})}{\delta v(\textbf{r}')}\bigg\rvert_{v_{0{\bf r}}} = \int d\textbf{r}' \chi^h_0\left(|\textbf{r}-\textbf{r}'|, v_{0{\bf r}}\right)\,,
\]
the approximate connector potential becomes a weighted average of the Kohn-Sham potential,
\begin{equation}\label{eqn:vc}
    v^c_{\bf r} = \ddfrac{\int d\textbf{r}' \chi^h_0\left(|\textbf{r}-\textbf{r}'|, v_{0{\bf r}}\right) v(\textbf{r}')}{\int d\textbf{r}' \chi^h_0\left(|\textbf{r}-\textbf{r}'|, v_{0{\bf r}}\right)} \,.
\end{equation}
We will call this linear response based approximation COT1 in the following.
Previous attempts exist to approximate an effective potential in the Thomas-Fermi expression using response theory \cite{Hilton1967,March1968,Lawes1980,Unger1971,Baldereschi1981}, instead of directly expanding the density, but to the best of our knowledge, these attempts have not been combined with modern \textit{ab initio} calculations.

In order to see how COT1 simulates higher orders, one may expand the resulting density $n^h(v^c_{\textbf{r}})$,

\begin{widetext}
  \begin{align}
   n^h(v^c_{\bf r})&=\frac{(-2v^c_{\bf r})^{3/2}}{3\pi^2}=\frac{1}{3\pi^2}\Big(-2v_{0{\bf r}}-2v_{0{\bf r}}\mathcal{V}\Big)^{3/2} 
   = \frac{1}{3\pi^2}(-2v_{0{\bf r}})^{3/2}\Bigg(1+\frac{3}{2}\mathcal{V}+\frac{3}{8}\mathcal{V}^2 + \dots \Bigg) \nonumber
    \\ &= n^h(v_{0{\bf r}})+\int d\textbf{r}' \chi^h_0\left(|\textbf{r}-\textbf{r}'|, v_{0{\bf r}}\right) \Delta v(\textbf{r}')+\frac{1}{6n^h(v_{0{\bf r}})}\Big(\int d\textbf{r}' \chi^h_0\left(|\textbf{r}-\textbf{r}'|, v_{0{\bf r}}\right) \Delta v(\textbf{r}')\Big)^2 + \dots\label{eq:byhands}
    \end{align}
\end{widetext}
where we defined \( \mathcal{V} = \ddfrac{1}{v_{0{\bf r}}}\ddfrac{\int d\textbf{r}' \chi^h_0\left(|\textbf{r}-\textbf{r}'|, v_{0{\bf r}}\right) \Delta v(\textbf{r}')}{\int d\textbf{r}' \chi^h_0\left(|\textbf{r}-\textbf{r}'|, v_{0{\bf r}}\right)} \) with \(\Delta v(\textbf{r}') = v(\textbf{r}')-v_{0{\bf r}}\), and where we used that $\int d\textbf{r}' \chi^h_0\left(|\textbf{r}-\textbf{r}'|, v_{0{\bf r}}\right)=\ddfrac{3}{2}\ddfrac{n^h(v_{0{\bf r}})}{v_{0{\bf r}}}$.
By construction, the COT result is exact to first order. However, also higher orders are added approximately.
If the system for which we wish to calculate the density is itself homogeneous, with a potential $v^h$ that is different from $v_{0{\bf r}}$, one finds from \eqref{eqn:vc} $v^c_{\bf r} = v^h$ and the resulting density is exact. The COT approximation for an inhomogeneous system can be understood order by order: for example, it replaces the second order response by a weighted product of first order responses. Again, this is exact in the limit of homogeneous $v({\bf r})$. 

COT results may be improved either by choosing a model that is closer to the real system, or by choosing a better approximation. Here, we keep the HEG as well known and simple model, and we will head for better approximations in the following.

\subsection{Optimal termination of a Taylor expansion}

\begin{figure}[h!]
    \centering
    \includegraphics[width=\columnwidth]{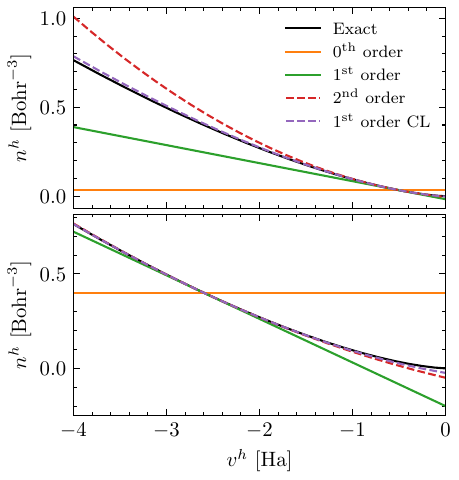}
    \caption{Taylor expansion of the density of a HEG, \(n^h(v^h)\), around \(v_0^h=-0.5\) Hartree (top panel) and around \(v_0^h=-2.5\) Hartree (bottom panel). The orange, green and red curves give the zeroth, first, second order expansions, respectively. The purple line is obtained by evaluating the derivative (the Lindhard function) at the midpoint, following \eqref{eq:taylor1} and the subsequent discussion in the text.}
    \label{result_taylor}
\end{figure}

One way to improve the result of low-order Taylor expansions is to follow the work of Schl\"omilch, Cauchy and Lagrange, who showed that a truncation at low order can be made exact if one changes the point where the derivatives are taken \cite{Cauchy1823,Lagrange1813}.
In particular, a first-order Taylor expansion expression
\begin{equation}
    f(x)=f(x_0) + (x-x_0)f'(\xi)
    \label{eq:taylor1}
\end{equation}
yields the exact result if the derivative $f'(x)$ is not necessarily taken at $x=x_0$, but at a point $\xi$ that lies between $x_0$ and $x$. For example, if the function $f(x)$ is a parabola, $f(x)=ax^2$ is exactly reproduced by \eqref{eq:taylor1} with the choice $\xi = \ddfrac{x_0}{2}+ \ddfrac{x}{2}$. As a consequence, any expandable function is correctly reproduced up to second order. The benefit with respect to a straightforward Taylor expansion can be seen easily by looking at the case of the density as function of a homogeneous potential $n^h(v^h)$, i.e., expression \eqref{eq:TF}. It is illustrated in Fig.~\ref{result_taylor}, where the expansion is shown for two different starting points. Taking the derivative at the midpoint significantly improves upon the first order. It is also better than a pure second-order expansion, because higher orders are not set to zero but taken into account approximately. This motivates us to move along this line.

The idea of taking the derivative at the midpoint to describe also the second order correctly, and higher orders approximately, can be generalized to functions of multiple variables, or vectors, and from the case of continuous variables, to functionals.
Applied to our problem, we may use the fact that  
\begin{equation}
    n(\textbf{r};[v]) \approx n(\textbf{r};[v_{0{\bf r}}]) + \int d\textbf{r}' \chi_0(|\textbf{r}-\textbf{r}'|;[v_{\xi}]) \Big( v(\textbf{r}') - v_{0{\bf r}}\Big)\,,
    \label{eq:taylor-cl}
\end{equation}
yields the exact result to second order, if 
\noindent\\
\( v_{\xi}(\tilde {\bf r})=v_m(\tilde {\bf r})\equiv \ddfrac{v_{0{\bf r}}}{2}+\ddfrac{v(\tilde {\bf r})}{2} \):
the ``midpoint'' potential $v_m$ is the inhomogeneous potential that is the average of the constant $v_{0{\bf r}}$ and of the actual, inhomogeneous potential $v$.
Following the same scheme as above, and using \eqref{eq:taylor-cl} as new approximation in \eqref{eq:cot-strategy}, both for the real density $n$ and for the HEG density $n^h$, leads to the new connector potential 
\begin{equation}\label{eqn:vc-cl}
   v^c_{\bf r} = \ddfrac{\int d\textbf{r}' \chi_0\left(\textbf{r},\textbf{r}'; [v_m]\right) v(\textbf{r}')}{\int d\textbf{r}' \chi^h_0\left(|\textbf{r}-\textbf{r}'|, \frac{v_{0{\bf r}}}{2}+\frac{v^c_{\bf r}}{2}\right)} \,.
\end{equation}

Note that because of the denominator on the right hand side of \eqref{eqn:vc-cl}, this is a self-consistent equation for $v^c_{\bf r}$ that can be solved, e.g., iteratively. An analytical solution is also possible, since \eqref{eqn:vc-cl} reduces to a cubic equation, noting that $\int d\textbf{r}' \chi^h_0\left(|\textbf{r}-\textbf{r}'|, v_{0}\right)=-\sqrt{-2v_0}/\pi^2$. However, we choose to solve for the connector iteratively to mitigate errors arising from our finite real-space grid. The problem, instead, is the numerator: it contains the general expression $\chi_0$, not the Lindhard function $\chi_0^h$, because the average $v_m$ of the potential $v$ of the real system and the homogeneous potential $v_{0{\bf r}}$ is itself inhomogeneous. This is of course a serious obstacle, and the approach as it is cannot be used for our purpose. However, the problem is similar to the initial one, where we had to find $n[v]$ for an inhomogeneous system: now we have to find $\chi_0[v_m]$. We can, therefore, again apply the COT strategy and, in analogy to \eqref{eq:cot-strategy}, ask for
\begin{equation}
    \chi_0({\bf r},{\bf r}';[v_m])=\chi_0^h(|\textbf{r}-\textbf{r}'|,v^c_{{\textbf r}{\textbf r}'})\,.
    \label{eq:chi0-connect}
\end{equation}
In other words, we ask for 
 $\chi_0({\bf r},{\bf r}';[v_m])$ evaluated for an inhomogeneous $v_m$ to equal the Lindhard function evaluated at some homogeneous connector potential $v^c_{{\textbf r}{\textbf r}'}$. This new connector potential will now in general be a different homogeneous potential for every pair ${\bf r}$ and ${\bf r}'$ in order to guarantee that the equality can be fulfilled. This is a crucial point: COT does not globally replace a material by a model, but it optimizes the model, in a generalized sense, locally. 

To find an approximation for $v^c_{{\textbf r}{\textbf r}'}$, we Taylor expand $\chi_0[v_m]$ and $\chi_0^h(v^c_{{\bf r}{\bf r}'})$ to first order around a homogeneous potential $v_{0{\bf r}{\bf r}'}$. Solving the resulting equation for $v^c_{{\textbf r}{\textbf r}'}$ yields  \begin{equation}
    v^c_{{\textbf r}{\textbf r}'}=\frac{v_{0\textbf{r}}}{2}
+\frac{\int d\textbf{r}_2\,\chi_0^{h,(2)}({\bf r}-{\bf r}_2, {\bf r}'-{\bf r}_2 ;v_{0{\bf r}{\bf r}'})v(\textbf{r}_2)}{2\int d\textbf{r}_2\,\chi_0^{h,(2)}({\bf r}-{\bf r}_2, {\bf r}'-{\bf r}_2 ;v_{0{\bf r}{\bf r}'})}\,,
\label{eq:vcrr-chi02}
\end{equation}
which is then to be used in \eqref{eq:chi0-connect}, transforming \eqref{eqn:vc-cl} into
\begin{equation}\label{eqn:vc-cl-con}
   v^c_{\bf r} = \ddfrac{\int d\textbf{r}' \chi^h_0\left(|\textbf{r}-\textbf{r}'|, v^c_{{\textbf r}{\textbf r}'}\right) v(\textbf{r}')}{\int d\textbf{r}' \chi^h_0\left(|\textbf{r}-\textbf{r}'|, \frac{v_{0{\bf r}}}{2}+\frac{v^c_{\bf r}}{2}\right)} \,.
\end{equation}
Maybe not surprisingly, now the second-order response function $\chi_0^{h,(2)}$ appears. One could in principle evaluate this expression using the non-interacting second order response function given, e.g., in \cite{Hammerberg1974,Pickenhain1976,Milchev1977,Witt2019}, but this would
increase the complexity and computational cost. Here, we will rather make a further approximation in order to explore how far one can get with simple expressions. Having $\chi_0^{h,(2)}$ in both numerator and denominator, thanks to the COT structure of the expression, has the advantage of potential error cancellation that may be further exploited in the future. For now, we therefore proceed with the most straightforward approximation, which is to suppose that the second-order response function $\chi_0^{(2)}$ is of short range with respect to both ${\bf r}-{\bf r}_2$ and ${\bf r}'-{\bf r}_2$, so that the dominant contributions arise from values of ${\bf r}_2$ located between ${\bf r}$ and ${\bf r}'$. This suggests the approximation to~\eqref{eq:vcrr-chi02},
 \begin{equation}
     v^c_{{\textbf r}{\textbf r}'}\approx \frac{v_{0\textbf{r}}}{2} + \frac{v(\lambda{\textbf r}' + (1-\lambda){\textbf r})}{2}\,,
     \label{eq:cl-app-gen}
 \end{equation}
 where $\lambda$ is a free parameter that can be optimized. We will call the connector approximation resulting from this potential average COT1-$\lambda$. A special case is $\lambda=1$, which leads to 
  \begin{equation}
     v^c_{{\textbf r}{\textbf r}'}\approx \frac{v_{0\textbf{r}}}{2} + \frac{v(\textbf{r}')}{2}\,:
     \label{eq:cl-app}
 \end{equation}
 this is a simplified midpoint approximation inspired by homogeneous systems that we will call COT1-av. It will be our starting hypothesis for the analysis of the approximations in the following. Instead, for $\lambda=0$ one has to replace $v(\textbf{r}')$ in \eqref{eq:cl-app} by $v(\textbf{r})$, which supposes increased importance of the local point where the density is calculated. As we will discuss below, $\lambda=1$ is indeed the most appropriate choice for helium, whereas small values of $\lambda$ are more appropriate for silicon and aluminum. 

The $\lambda$-approximation to the second-order response function in \eqref{eq:vcrr-chi02} is only applied to the inhomogeneous system in the numerator, not to the denominator, which stems from the HEG. Such an approximation therefore introduces an imbalance: for the HEG, the modified first-order expansion yields a result that is exact to second order, but not for the real system, unless it is itself homogeneous. On the other hand, $\lambda$ is a parameter that can be optimized. Only at this level of approximation, instead of straightforward mathematics, some engineering is used.

\eqref{eq:cl-app-gen} is one of the simplest ways to do so; another possibility to mix ${\bf r}$- and ${\bf r}'$-dependence is to introduce an ${\bf r}$-dependent weight $\alpha_{\bf r}$, replacing \eqref{eq:vcrr-chi02} by 

\begin{equation}
    v^c_{{\textbf r}{\textbf r}'}\approx \frac{v_{0\textbf{r}}}{2}+\alpha_{\bf r}v(\lambda{\textbf r}' + (1-\lambda){\textbf r})\,.
    \label{eq:vcrr-app-alpha}
\end{equation}

These two cases (namely $\lambda$ in \eqref{eq:cl-app-gen} and $\alpha_\mathbf{r}$ in \eqref{eq:vcrr-app-alpha}) will constitute the only free parameters in our approximations, and they will be optimized in the following section. Table \ref{tab:all} summarizes the main approximations and their acronyms used in this work.
\begin{table}[bh]
    \centering
        \caption{Summary of approximations used in this work. Expressions for the connector are then used in \eqref{eq:cot-strategy}.}
    \begin{tabular}{ c c}
        \hline \hline
        Method & Approximation\\
        \hline \hline
        LPA &  \eqref{eq:TF} with $v_{0{\bf r}}=v({\bf r})$ \\  
        COT1 & \eqref{eqn:vc} with $v_{0{\bf r}}=v({\bf r})$\\  
        COT1-av &  \eqref{eqn:vc-cl-con} \& \eqref{eq:cl-app} \\  
        COT1-$\lambda$ &  \eqref{eqn:vc-cl-con} \& \eqref{eq:cl-app-gen} \\  
        COT1-$\alpha$ &  \eqref{eqn:vc-cl-con} \& \eqref{eq:vcrr-app-alpha} \\  
        \multirow{2}{*}{COT1-av-KS$^*$} & \eqref{eqn:vc-cl-con} \& \eqref{eq:cl-app} \\ & in a SC-KS loop \\ \hline \hline
    \end{tabular}
    \label{tab:all}
\end{table}

\section{Results}
\label{sec:results}

In order to explore the performance of the various approximations, we will study a very inhomogeneous system: cubic helium \cite{Schuch1961}. Since our approximate expressions become exact when the system is homogeneous, this is a severe test. 
We will also include systems with more slowly varying density, namely, silicon and aluminum. 
For all systems, the input potential is the KS potential obtained from a fully converged Kohn-Sham cycle using a local norm-conserving pseudopotential and the LDA \footnote{Since there is no local pseudopotential for silicon within standard KS-DFT, we will use a local pseudopotential developed for orbital-free DFT \cite{Zhou2004,XU2024}}. The resulting densities are our benchmark.
Computational details can be found in Appendix \ref{sec:compdet}.

\subsection{General behavior of the approximations}
\label{subsec:results-general}

\begin{figure}
    \centering
    \includegraphics[width=\columnwidth]{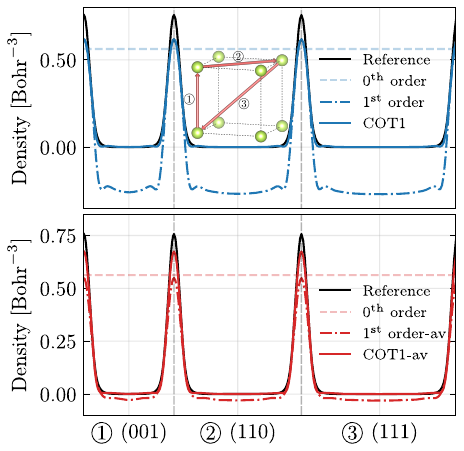}
    \caption{Charge density of cubic helium resulting from a given LDA KS potential. In black, benchmark result from the solution of the KS equation. All other curves are approximations. The horizontal dashed line is the zero-order result, using a homogeneous potential $v_{0{\bf r}}$ that is the average potential of the system.    
    Upper panel: blue dot-dashed, linear response. Blue continuous line: COT1 result, where the linear response is used to approximate the connector potential, \eqref{eqn:vc}.
    Lower panel: red dot-dashed, linear response with derivative taken at the average of two homogeneous potentials to approximate the approach of Cauchy and Lagrange, \eqref{eq:cl-app} used in \eqref{eq:chi0-connect}. Red continuous, this improved linear response-like approximation is used to build a better connector, COT1-av, \eqref{eqn:vc-cl-con}.}
    \label{fig:check}
\end{figure}

Let us first illustrate the upsides and downsides of the first-order Taylor expansion, as well as the benefit of COT, at the example of helium. For simplicity, the starting $v_{0{\bf r}}$ of the expansion is chosen to be the same everywhere and equal to the average potential. The upper panel of Fig.~\ref{fig:check} shows the zero-order result, i.e., the HEG density of this flat potential, as well as the first-order result \eqref{eq:taylor-1} and the COT1 result $n^h(v^c_{\bf r})$ with $v^c_{\bf r}$ from \eqref{eqn:vc}. These results are compared to the benchmark density from the KS calculation along a high-symmetry path in the unit cell, as indicated in the figure. 

The first-order expansion performs astonishingly well in view of the fact that the starting $v_{0{\bf r}}$, and therefore also the zero-order density, are so different from the real KS potential and density. The result on and around the atoms is qualitatively reproduced, although the maximum is too low. However, the first-order density is unphysical in the low-density regions, since the first-order correction overshoots, which leads to negative densities. This flaw is completely cured by the application of COT: while the COT1 result is very close to the already reasonable first order in the high-density regions, there is significant improvement in the regions of lower densities. Indeed, COT automatically fulfills the exact constraint of positive density, because it yields a HEG density at every point.  
It implicitly adds higher-order corrections, as highlighted in equation \eqref{eq:byhands}, which leads to overall much better results without any significant additional computational cost, compared to the linear-response approximation. 

The lower panel of Fig.~\ref{fig:check} makes use of the simplest approximate Cauchy-Lagrange correction to the first order, \eqref{eq:cl-app} used in \eqref{eq:chi0-connect}. This correction greatly improves the linear response in the low density regime, but the density is still slightly negative, and the maximum of the density on the atoms is even more underestimated than in the pure linear response. To explain the local degradation of the result on the atoms, one should remember that on top of choosing the derivative such that only the second order, and not all orders, are in principle reproduced exactly, we also approximate \eqref{eq:vcrr-chi02} by \eqref{eq:cl-app}. This approximation is expected to be particularly severe in regions of quickly varying density. However, again the problems are cured by using this approximately corrected linear response as input for COT: The COT1-av density is throughout positive, the maximum is closer to the benchmark compared to all other approximations, and we also find improvement at the bottom of the atoms, where the connector potential has to overcome the wiggles of the KS potential (see Fig.~\ref{connectors}).

\subsection{Starting from the local potential approximation}
\label{subsec:results-lpa-start}
In practice, one would of course try to optimize the starting point of an expansion. Note again that we always expand around a homogeneous system, but we may take a different homogeneous starting potential $v_{0{\bf r}}$ at every point in space. Indeed, for an inhomogeneous system such as helium, we may expect that this freedom is important.

\subsubsection{Local Potential Approximation}
\begin{figure}[t]
    \centering
    \includegraphics[width=\columnwidth]{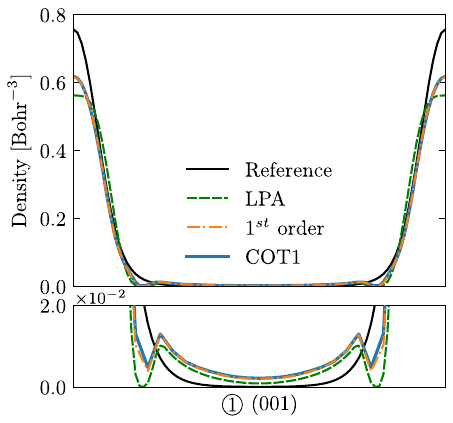}
    \caption{Charge density of cubic helium. In black, benchmark result from the solution of the KS equation. All other curves are approximations starting from the LPA, which is the green dashed curve. A linear expansion around the LPA yields the orange dash-dotted line. Blue continuous line is the COT1 result, where the linear response is used to approximate the connector potential, \eqref{eqn:vc}. The lower panel zooms into the low-density region.}
    \label{fig:local}
\end{figure}

A natural choice is $v_{0{\bf r}}=v(\textbf{r})$, the local KS potential. This choice is motivated by the nearsightedness principle introduced by Walter Kohn \cite{Kohn1996,Prodan2005}, which states that electronic properties at a given point are primarily influenced by the nearby environment. 
For our case, where we search for the density as a functional of the potential, the approximation
\begin{equation}
    n({\bf r};[v])\approx n^h(v_{0{\bf r}}=v({\bf r}) )
    \label{eq:lpa}
\end{equation} 
would become exact if the variations of $v({\bf r})$ were much slower than the decay range of the linear and all higher order response functions. We will call \eqref{eq:lpa}, which is also directly obtained from the approach of Thomas and Fermi \cite{Thomas1927,Fermi1928} and which has already been used as an approximation for a potential-functional \cite{Gross2009}, the Local Potential Approximation (LPA). 
It is the potential-functional analogue to the LDA, where the KS exchange-correlation potential is determined as the HEG result taken at the local density; put simply, here the roles of the potential and of the density are swapped. Indeed, in spite of this analogy, it should be stressed that the LDA and the LPA address two fundamentally different problems. The LDA aims at importing interaction effects from the HEG into the real system, and it acts as an approximation to an auxiliary potential. The LPA aims at importing from the HEG the effect of the Schr\"odinger equation, i.e., of the (non-local) kinetic energy operator, and it directly approximates the observable of interest, the density. 
These differences could lead to a very different degree of nearsightedness. Moreover, here we try to approximate directly an observable rather than some auxiliary potential, which will require a higher accuracy, and it is therefore more ambitious.
One major aim of the present work is indeed to see how far one can go with a functional approximation directly for observables.

\begin{figure}[t]
    \centering
    \includegraphics[width=\columnwidth]{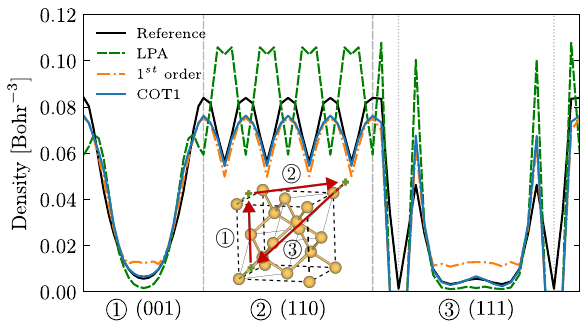}
    \caption{Charge density of silicon. In black, benchmark result from the solution of the KS equation. All other curves are approximations starting from the LPA, which is the green dashed curve. A linear expansion around the LPA yields the orange dashdotted line. Blue solid line is the COT1 result, where the linear response is used to approximate the connector potential, \eqref{eqn:vc}. The trajectory starts at the center between the two Si atoms, and goes through the $(001)$, $(110)$ and $(111)$ directions, as indicated in the inset. The two Si atoms along the path are indicated by the vertical dotted lines.}
    \label{fig:si1}
\end{figure}

Figures~\ref{fig:local} and~\ref{fig:si1} show the performance of the LPA for cubic helium, and for silicon, respectively. Remarkably, despite its simplicity and negligible computational cost, LPA already captures the general shape of the density of cubic helium, which implies that the latter essentially follows the structure of the Kohn--Sham potential \( v(\mathbf{r}) \). It is of course known that this Thomas-Fermi approximation is in general not sufficiently accurate, see, e.g., \cite{Gross2009,Elliott2008,Cangi2010,Cangi2011,CANGI2013,Mi2023,March1957,Lieb1977,Parr1994}.
Also, for cubic helium the result would not be good enough for most applications: the LPA underestimates the density on the atoms and overestimates it in the low-density region, a zoom of which is shown in the lower panel. The LPA in Fig.~\ref{fig:local} also shows the wiggles at the bottom of the atoms that mirror the behavior of the KS potential, but that are not seen in the benchmark density. Moreover, the number of electrons integrated over the unit cell is $N_\mathrm{el}^\mathrm{LPA} = 2.26$, which is much too high. In the case of silicon in Fig.~\ref{fig:si1}, the LPA by definition still mirrors the KS potential. This leads to a density that is qualitatively wrong in some places: in particular, the density oscillation in the (110) direction has opposite phase with respect to the benchmark result. Moreover, the density is too low at the covalent-bond center (i.e., at the peaks
along the bond path along (001)). Instead, the number of electrons integrated over the unit cell is exactly constrained thanks to the freedom in aligning the chemical potentials of the real system and the model (see Appendix \ref{app:mu}). The LPA can be seen as an approximation to COT, with $v^c_{\bf r}\approx v({\bf r})$, i.e., a COT approximation of perfect nearsightedness. In the following, we will use the LPA merely as a convenient starting point for the subsequent expansions.

\subsubsection{Linear response and connector starting from LPA}
Figure~\ref{fig:local} and~\ref{fig:si1} show that the first-order correction brings the density on the atoms and its decay closer to the benchmark result, compared to this already significantly improved starting point. Moreover, the density is always positive, as opposed to Fig.~\ref{fig:check}, for both systems. For silicon, the linear response (in orange) is able to correct the qualitatively wrong behavior of LPA along the (110) direction. Moreover, COT1 improves on top of this by further correcting the density in the low-density region. This suggests that the higher-order corrections are properly simulated by COT1. Yet, for the case of cubic helium, the results of linear response are too similar to the LPA to be useful in practice, and COT does not add any significant correction
with respect to this outcome. This can be rationalized by looking at the linear-response expression \eqref{eq:taylor-1}, which is also the basis of COT. Linear response cannot yield a significant contribution if $\chi_0(|{\bf r}-{\bf r}'|;[v_{0{\bf r}}])$ decays rapidly as a function of $|{\bf r}-{\bf r}'|$. Therefore, when $v_{0{\bf r}}=v({\bf r})$, the low-density region suffers from a too short ranged response function when the starting point is the LPA. This is particularly severe in cubic helium, where the density (thus the absolute value of the local potential) is very low in the interstitial region. An illustration can be found in Fig.~\ref{fig:response} in the Appendix. To overcome this problem, higher orders must be included at least approximately in a way that is different from the COT1: using \eqref{eq:byhands} for analysis of the connector result, it appears that the smallness of the first-order contribution propagates to all higher orders, which makes COT1 unable to solve the problem of the first-order result. 

\subsubsection{Improved termination of perturbation series}\label{subsec:improved_conn}
\begin{figure}[t]
    \centering
    \includegraphics[width=\columnwidth]{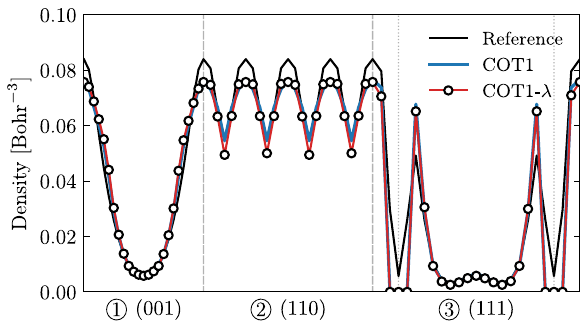}
    \caption{Charge density of silicon calculated using \eqref{eq:cl-app-gen} with $\lambda=0.1$ in comparison to COT1 (blue) and the reference density in black.}
    \label{fig:si2}
\end{figure}

We start by considering silicon. Since COT1 yields a relatively good charge density (we will quantify what is meant by good below), we can directly tackle \eqref{eq:vcrr-chi02} and see whether we can further improve the results by a better approximation of the second-order response function. For this, we use the COT1-$\lambda$ approximation, using the value of $\lambda$ that minimizes the Mean Absolute Difference Error (MADE)
 of the density $\epsilon^\mathrm{MADE}_n$, see Sec. \ref{subsec:quant} below, calculated over the unit cell. The resulting charge density is shown in Fig.~\ref{fig:si2} for the optimal value of $\lambda=0.1$. The only visible difference with respect to the already very good COT1 result is in the (110) direction, where the width of the density oscillation is slightly improved, though still underestimated. This seems to suggest that there is no quantitative improvement in the overall accuracy of the charge density, but we will see in Sec. \ref{subsec:observables} that the observables calculated from COT1-$\lambda$ are significantly improved. 
 
\begin{figure}[t]
    \centering
    \includegraphics[width=\columnwidth]{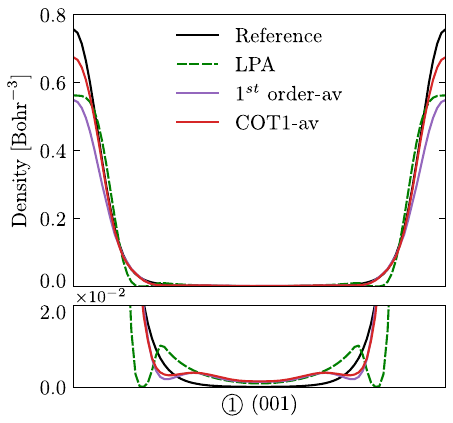}
    \caption{Charge density of cubic helium. In black, benchmark result from the solution of the KS equation. All other curves are approximations, starting from the LPA, which is the green dashed curve. The violet curve is the linear response with derivative taken at the average of two homogeneous potentials to approximate the remainder of Cauchy and Lagrange, \eqref{eq:cl-app} used in \eqref{eq:chi0-connect}. Red continuous, this improved linear response-like approximation is used to build a better connector, \eqref{eqn:vc-cl-con}. The lower panel zooms into the low-density region.}
    \label{fig:sc-bilocal}
\end{figure}

 The case of cubic helium is more interesting, since linear response on top of the LPA, and therefore COT1, are not powerful enough (especially in the low density regions). In this case, the optimal value is $\lambda=1$, i.e., the intuitive approximation \eqref{eq:cl-app}. This makes sense, since we have already seen that a more non-local response function is needed in the low-density region for cubic helium. Let us first examine the improved Taylor expansion itself, i.e., the linear response \eqref{eq:taylor-cl} with the approximate Cauchy-Lagrange correction, \eqref{eq:chi0-connect} with \eqref{eq:cl-app}. The result is shown in Fig.~\ref{fig:sc-bilocal}. As in Fig.~\ref{fig:check} and in comparison to Fig.~\ref{fig:local}, we find that the Cauchy-Lagrange correction improves the results in the low density range, while it further lowers the already too low density on the atoms. 
Instead, again similarly to Fig.~\ref{fig:check}, using now COT on top of this approximation leads to the COT1-av result, which is closer to the benchmark than all other approximations for the whole range of densities.

Of course, we still use the quite rough approximation of the second-order response function that uses \eqref{eq:cl-app} instead of \eqref{eq:vcrr-chi02}.  We will therefore explore whether another still simple and pragmatic approximation to \eqref{eq:vcrr-chi02} could lead to significant improvement, following \eqref{eq:vcrr-app-alpha}. For this, we have to make an ansatz for the factor $\alpha_{\bf r}$ with a few parameters that can be determined approximately. In this work, we will address the questions of how many parameters are needed in order to obtain a significantly increased accuracy, and to what extent these parameters are transferable.

\begin{figure}[t]
    \centering
    \includegraphics[width=\columnwidth]{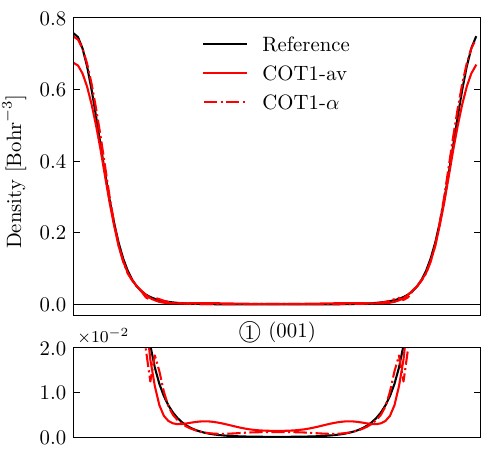}
    \caption{Charge density of helium calculated using \eqref{eq:vcrr-app-alpha} with \eqref{eqn:alpha_opt} and the parameters $A$ and $B$ of $\alpha_\textbf{r}=A(n^{\rm LPA}({\bf r}))^B$ taken from a fit to the benchmark Kohn-Sham calculation. In black, the reference density. Red dash-dotted line, result of the fit. Red solid line, the most advanced parameter-free result, which is COT1-av, \eqref{eqn:vc-cl-con} with \eqref{eq:cl-app}. Lower panel: zoom on the low-density region.}
    \label{fig:alpha_opt_uncompressed}
\end{figure}

The idea is that $\alpha_{\bf r}$ should bring in an ${\bf r}$-dependence that, using the nearsightedness principle, is directly linked to a local physical descriptor: this could be either the potential or the density itself. In the case of the potential, this is equivalent to asking for a local dependence on the density determined by the LPA, since $n^\mathrm{LPA}$ is directly related to the local potential, i.e. $n^\mathrm{LPA}({\bf r}) \propto v^{(3/2)}(\textbf{r})$. One simple ansatz is  
\begin{equation}\label{eqn:alpha_opt}
   \alpha_\mathbf{r}=A\big(n(\mathbf{r})\big)^B\,.
\end{equation}
To test this ansatz and to compare the two options, for simplicity we use $n=n^{\rm ref}$ and $n=n^\mathrm{LPA}$, the benchmark and the LPA density, respectively. The former will of course be unknown in practice, but we may expect a similar behavior when $n^{\rm ref}$ is replaced by any approximation that is significantly better than the LPA. Figure~\ref{fig:alpha_opt_uncompressed} shows the results obtained by fitting the two parameters for the option $n^{\rm LPA}$ to the benchmark result along the $(001)$ path shown in the figure. The two options give almost indistinguishable results, and they allow us to obtain excellent agreement to the benchmark curve, with only the two parameters and despite the simplicity of the expression. This is not only true along the path shown in Fig.~\ref{fig:alpha_opt_uncompressed}, where the parameters have been fitted, but it also holds throughout the unit cell. This is a first indication for the transferability of these parameters. We find $A^{\rm ref} = 0.6773$ and $B^{\rm ref} =0.1455$ when $n=n^{\rm ref}$ and $A^{\rm LPA} = 0.7165$ and $B^{\rm LPA} =0.1919$ when $n=n^{\rm LPA}$. This is to be compared with approximation \eqref{eq:cl-app} that corresponds to $A=0.5$ and $B=0$. The two parameters are very efficient: there is substantial improvement both in the high density and in the low density region.

The fit of the parameters to the benchmark result does, of course, not tell us how much the improvement is due to a better description of $\chi_0^{(2)}$, and how much it is due to an implicit correction of higher orders. This will be important when one tries to develop more advanced parameter-free expressions. For now, the most important question is the transferability of the parameters: if the fit is to be of more than purely intellectual interest, one must be able to perform the fit for one system and use the resulting parameters in, at least slightly, different systems. In the following, we will check our approximations, including this transferability of parameters, by substantially compressing and expanding the system.

\subsubsection{Compressed and expanded helium}

\begin{figure}
    \centering
        \includegraphics[width=\columnwidth]{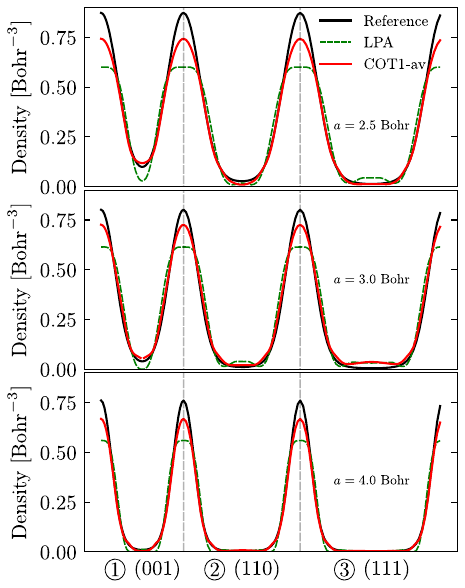}
    \caption{Charge density of cubic helium for 3 different compressions. From top to bottom, lattice constants are $a=2.5,3.0,4.0$ Bohr. In black, benchmark result from the solution of the KS equation. The other two curves are approximations, starting from the LPA, which is the green dashed curve. Red solid line, the most advanced parameter-free result, which is COT1-av, \eqref{eqn:vc-cl-con} with \eqref{eq:cl-app}.}
    \label{fig:compressed}
\end{figure}

The above results for helium were obtained for a lattice constant of $a=8.016$ Bohr (see Appendix \ref{sec:compdet}). Our aim in the following is to study strong compressions or expansions that truly change the charge distribution by creating or suppressing hybridization between the atoms. The three panels in Fig.~\ref{fig:compressed}, which are also to be compared to Fig.~\ref{fig:sc-bilocal}, show the reference density, the LPA and the COT1-av results for $a=2.5$, $a=3.0$ and $a=4$ Bohr. 
For the stronger compressions, especially in the (001) direction, there is no longer a region of very low density (note the change of scale of the horizontal axis that is normalized to the unit cell). The evolution is not a trivial overlap of atoms. It is therefore not clear whether the most compressed system will behave like a more homogeneous or like a less homogeneous system. The LPA results, which should be perfect for slowly varying density, seem to have a slight tendency to perform less well as the system is compressed; this will be analyzed further below, in Subsec.~\ref{subsec:quant}. 
The COT1-av corrects a substantial part of the error in all cases; it erases the typical wiggles of the LPA at the bottom of the atoms, and it is significantly better in most of the low density regions. Also the rise of the density close to the atoms is very well described, until a point where the LPA flattens and therefore strongly underestimates the maximum density on the atoms. The COT1-av curve does not flatten, although it still departs from the reference curve and leads to a too low maximum density. 

\begin{figure}[h!]
    \centering
    \includegraphics[width=\columnwidth]{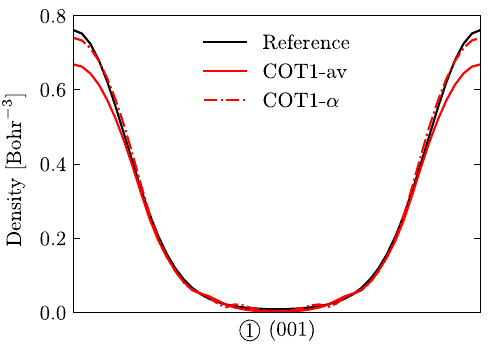}
        \includegraphics[width=\columnwidth]{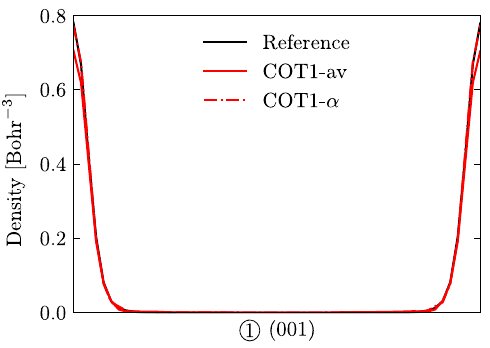}
    \caption{Charge density of helium at different lattice constants. Upper panel: lattice constant $a=4$ Bohr. Lower panel: lattice constant $a=15$ Bohr. In black, the reference density. In solid, COT1-av result. Dash-dotted, COT1-$\alpha$ approximation using $\alpha_\textbf{r}$ with parameters taken from a fit to a Kohn-Sham calculation at $a=8.016$ Bohr.}
    \label{alpha_opt_L4}
\end{figure}

Finally, varying the lattice constant also allows us to test the transferability of the parameters A and B in $\alpha_{\bf r}$. The upper panel of Fig.~\ref{alpha_opt_L4} shows the charge density for $a=4 $ Bohr calculated using $\alpha_{\bf r}$ with the parameters obtained from a fit to the results shown in Fig.~\ref{fig:alpha_opt_uncompressed} and obtained with $a=8.016$ Bohr. While very small deviations from the reference results can be detected, in particular on top of the atoms, the agreement is still impressive. The lower panel of Fig.~\ref{alpha_opt_L4} shows results for the expanded solid, with $a=15 $ Bohr. COT1-av performs well everywhere, except for an underestimation of the density on top of the atoms. The result of COT1-$\alpha$ with the parameters of $\alpha_{\bf r}$ fitted at $a=8.016$ Bohr is in excellent agreement with the benchmark. Altogether, these results illustrate the capability of the connector approach to significantly improve results of a Taylor expansion for comparable computational cost, and the possibility to obtain further substantial improvement by introducing a few transferable parameters. In the next subsection, we will quantify these statements. Since the case of cubic helium proves to be the most challenging due to the very low density in the interstitial region, we will focus more in detail on this system.

\subsection{Error quantification}
\label{subsec:quant}
\begin{figure}
    \centering
    \includegraphics[width=\columnwidth]{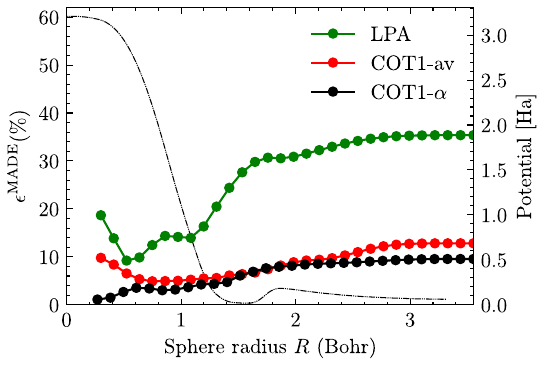}
    \includegraphics[width=\columnwidth]{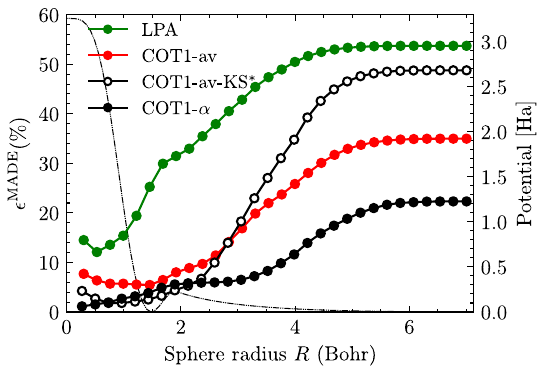}    
    \includegraphics[width=\columnwidth]{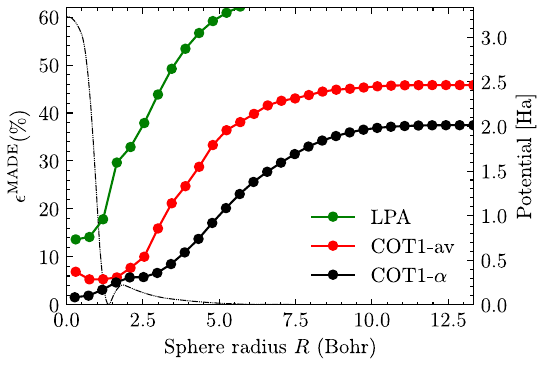}
    \caption{MADE error on the density of cubic helium defined on a sphere with radius $R$ centered on an atom as a function of $R$, \eqref{eq:made}. Upper panel: compressed structure with $a=4.0$ Bohr. Middle panel: equilibrium geometry. Lower panel: expanded structure with $a=15.0$ Bohr. In green, LPA results. In red, COT1-av. In black, COT1-$\alpha$. Open black circles in the middle panel, denoted COT1-av-KS$^*$ are obtained using COT1-av in a self-consistent Kohn-Sham cycle. The black dot-dashed line is the absolute value of the potential, with its scale on the right vertical axis.}
    \label{fig:made_accumulation_uncompressed} 
\end{figure}

All the above comparisons are mainly qualitative. Moreover, the impact of errors will depend on what one wants to do with the density. Therefore, Fig.~\ref{fig:made_accumulation_uncompressed} shows 
the MADE defined on a sphere with radius $R$ centered on a helium atom, as a function of $R$:
\begin{equation}
    \epsilon^\mathrm{MADE}_n(R) = \ddfrac{\int_{|\mathbf{r}|\leq R} \, d\mathbf{r} \left|n^\mathrm{ref}(\mathbf{r})-n(\mathbf{r})\right|}
    {\int_{|\mathbf{r}|\leq R} d\mathbf{r}\, n^\mathrm{ref}(\mathbf{r})} \times 100\,.
    \label{eq:made}
\end{equation}
For small $R$ this error is dominated by the high density regions, whereas large {$R$} take into account mostly the low-density regions; note that most of the unit cell consists of regions of very low density. The results for a compressed, for the equilibrium and for the expanded structure are shown in the upper, middle and lower panel, respectively. For analysis, the figure also shows the absolute value of the potential in each case; in order to appreciate its width, note the change of scale of the horizontal axis.

The LPA starts with a sizeable error on the atom that is the largest for the compressed structure, where the atomic potential is the sharpest. The error then shows a short decrease before it increases again when the density becomes steeper, with another rise where the potential shows a wiggle at the bottom of the atom. For large $R$ the error picks up mostly the low-density region, where it reaches a plateau. This plateau is the lowest for the compressed structure, which reflects the effect of a smoother potential upon compression: as expected, the LPA performs best for slowly varying densities. 

The most advanced parameter-free approximation, COT1-av, yields a much better result: the error on the atom is now of the order of 10\% or lower in all cases, and it decreases to about 5\% for increasing $R$. It increases again towards a plateau when $R$ is in a region of significantly lower density, with an error of the order of 10\% in the compressed case, and about 35 -- 45\% in equilibrium and in the expanded case. 
Again,
these results are consistent with the expectation that our approximation works best when the density is more slowly varying. The helium results also illustrate the limitation of our linear-expansion-based approach in the low-density region, where the response function has a tendency to be too short ranged to yield an efficient correction, and where, of course, small absolute errors yield large contributions to the MADE. Future optimization may take this observation into account by using a starting point that is different from the local potential in regions of very low density.

Finally, COT1-$\alpha$ benefits from the two fit parameters of the method. It is therefore not surprising that the result is improved for the equilibrium case, where the parameters have been fitted: there is almost no error on the atom and less than 5\% in a region where the density is significant, which means that in that region our approximation reaches the typical accuracy of the Local Density Approximation that has been used to calculate our reference density \cite{Aouina2023, Chen2021}. For larger $R$ the error increases again: indeed, the fit of $A$ and $B$ was done along a line in the (001) direction, which gives less weight to the low-density regime than the 3D MADE, and it is therefore a simple, but not optimized fit for the whole unit cell. When transferred to the compressed and to the expanded structures, the same parameters $A$ and $B$ are still very beneficial: the error on the atom remains vanishingly small, and for no value of $R$ does it exceed the error of COT1-av. The smallest correction is found on the strong wiggle of the potential. Furthermore, $\alpha_{\bf r}$ substantially improves the already good result where the density is significant, and it also systematically lowers the error plateau. This turns COT1, augmented with a simple ansatz containing a few parameters, into a promising route for future developments.

Since silicon does not present the same low-density challenges as helium, we do not show a figure similar to Fig.~\ref{fig:made_accumulation_uncompressed}. Instead, in Table \ref{tab:si_made} we report the MADE errors calculated over the unit cell, which correspond to $R\to \infty$ in \eqref{eq:made}. To show their robustness, in Table~\ref{tab:si_made} we also include results for aluminum. For both materials, the LPA performs poorly, and each improvement systematically reduces the MADE error. A crucial finding is that the optimal $\lambda$ for both silicon and aluminum are around the same value (see Appendix~\ref{sec:app_lambda}), and chosen to be $\lambda=0.1$ in this work. This is in stark contrast to helium (and its compressed and expanded versions), where the optimal $\lambda$ is significantly different. Instead, COT1-$\alpha$ results shown below are obtained for all materials with the $\alpha_{\bf r}$ parameters optimized for cubic helium discussed in Subsec.~\ref{subsec:results-lpa-start}. Table~\ref{tab:si_made} shows that at every step, the error decreases. It is noteworthy that this improvement is obtained by using the same two parameters in $\alpha_{\bf r}$ for helium, silicon, and aluminum. This suggests that while the optimal $\lambda$ may vary depending on the class of materials, the underlying physics captured by $\alpha_{\bf r}$ remains robust and transferable across different systems, leading to the most successful method for decreasing the MADE.

\begin{table}[bh]
    \caption{\label{tab:si_made}%
    Mean absolute difference errors $\epsilon^{\mathrm{MADE}}$ (\%), integrated
    over the whole unit cell, of different approximations for the density of silicon
    and aluminum. The number of electrons is exactly constrained for all methods;
    $\lambda=0.1$ for both silicon and aluminum, and the $\alpha_{\mathbf{r}}$
    parameters are the same for silicon, aluminum, and cubic helium.}
    \begin{tabular}{l S[table-format=2.2] S[table-format=2.2]}
    \hline \hline
    Method & {Si (\%)} & {Al (\%)} \\
    \hline
    \addlinespace[3pt]
    LPA            & 33.37 & 35.99 \\
    LRA            & 12.54 & 8.87 \\
    COT1           & 8.96 & 9.12 \\
    COT1-$\lambda$ & 8.41 & 8.27 \\
    COT1-$\alpha$  & 7.88 & 7.48 \\
    \hline \hline
    \end{tabular}
\end{table}

\subsection{Observables derived from the density}
\label{subsec:observables}

\begin{figure}
    \centering
    \includegraphics[width=0.9\columnwidth]{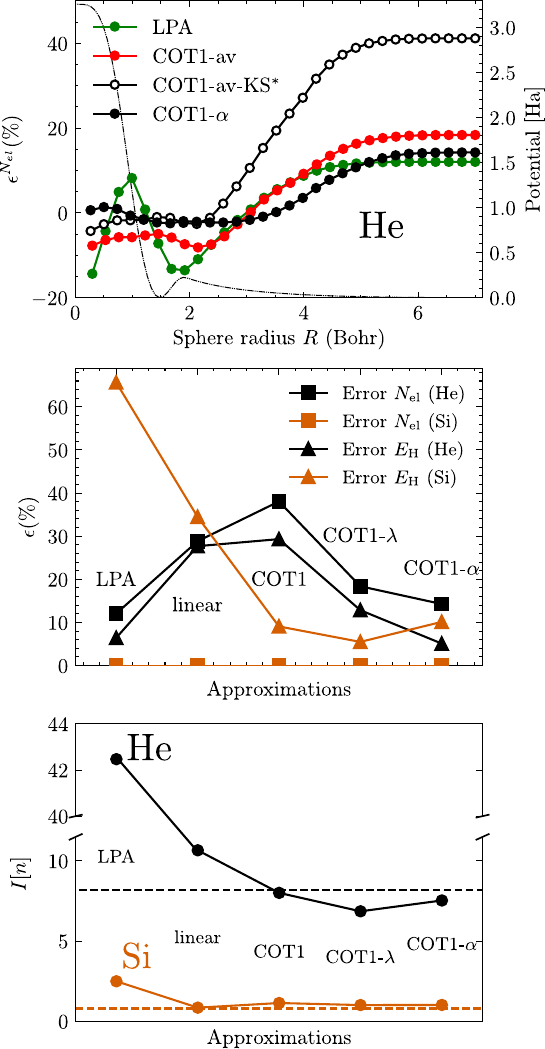}
    \caption{Errors in quantities calculated with the density, for various approximations. Upper panel: signed error on electron number integrated on a sphere for cubic helium (see text). The black dot-dashed line is the absolute value of the potential, with its scale on the right vertical axis. Middle panel: error on Hartree energy (triangles) and on absolute electron number (squares) for He and Si. Bottom panel: information functional for He and Si (see text). The black and orange dashed line represents the value of the information functional evaluated for the reference densities of He and Si, respectively.}\label{fig:observables}
\end{figure}

With the mean error being very different in different regions, it is important to investigate its impact on observables derived from the density. The most immediate question is the number of electrons. Figure \ref{fig:observables} compares errors in different observables calculated with the density that was determined using the various approximations; the results are obtained in the equilibrium structure. The top panel, dedicated to helium, gives the percentage error of the number of electrons in a sphere as a function of sphere size $R$, defined as $\epsilon^{N_{\rm el}}=
\dfrac{
\int_{|\mathbf{r}|\leq R}  \,d\mathbf{r}\,
(n^\mathrm{ref}(\mathbf{r})-n(\mathbf{r}))}
{\int_{|\mathbf{r}|\leq R} \,d\mathbf{r}\,
n^\mathrm{ref}(\mathbf{r}) 
}
\times 100\,.$ LPA leads to large negative and positive errors, reflecting both the wrong distribution and the wrong total number of electrons, with $N^{\rm LPA}_{\rm el} = 2.26$ instead of $N_{\rm el}=2$. COT1-av has milder fluctuations, but it is slightly electron-deficient on the atoms, and the overly high density in the low-density regions leads to an overall number of electrons that is too high, with $N^{\rm COT1-av}_{\rm el} = 2.37$. COT1-$\alpha$ performs better everywhere, although it still overestimates the lower densities, leading to $N^{\rm COT1-{\alpha}}_{\rm el}=2.29$ \footnote{Interestingly, \cite{Clay2026} also found that Thomas-Fermi based approximations for the energy give improved results when one allows for too large number of electrons}. We do not report the same figure for silicon, since the number of electrons is exactly constrained in all approximations.

The middle panel shows the percentage error of the Hartree energy. On the same panel, we also indicate the percentage error of the electron number for the different approximations (which is zero for silicon). Interestingly, the two errors clearly correlate for cubic helium, although the divergent contribution to the Hartree energy, which stems from the average density, is omitted as usual. In helium, while the LPA has the largest MADE on the density, as shown in Fig.~\ref{fig:made_accumulation_uncompressed}, it performs relatively well on the Hartree energy, better than COT1-av and only slightly better than COT1-$\alpha$. This is most likely due to error cancellation: already the sphere-integrated electron number in the upper panels shows that the LPA has comparably large positive and negative errors in the region of the atom, whereas the COT1-av error in that region is smaller, but always negative. The case of silicon, instead, shows a clear trend of improvement due to COT, with a reduction of the error by more than a factor of 5 starting from 65.7\% for the LPA.

The lower panel shows the result for an information functional $I[n] =
\dfrac{\int \frac{|\nabla n(\mathbf r)|^2}{n(\mathbf r)} \, d\mathbf r}
{\int n(\mathbf r)\, d\mathbf r}$, which measures the localization of the electronic density \cite{NAGY2025}. This functional is directly proportional to the von Weizs\"{a}cker kinetic energy~\cite{TFvW-functional}: $I[n] = 8\,T_\mathrm{vW}[n]/N$, linking the localization measure to a kinetic energy functional
\footnote{Equivalently, writing \( \varphi = \sqrt{n} \), one obtains \( I[n] = \frac{4}{N}\|\nabla\varphi\|_{L^2}^2 \), i.e., \(I[n] \) is proportional to the square of the $H^1$ seminorm of \(\sqrt{n} \). This connects to a rigorous result in DFT: the domain on which the Lieb universal functional \( F[n] \) is finite is precisely \( \mathcal{A}_N = \{ n \geq 0 : \sqrt{n} \in H^1(\mathbb{R}^3),\, \int n(\mathbf r) \,d\mathbf{r} = N \} \)~\cite{Lieb1983, Corso2026}. The \(H^1 \) condition on \( \sqrt{n} \) is physically equivalent to requiring a finite von Weizsäcker kinetic energy. Convergence in this space controls gradient-dependent quantities, making \( I[n] \) a mathematically principled and physically motivated quality measure for approximate densities, beyond what \( L^2\) or pointwise errors can capture}. 
Unlike the mean absolute deviation of the density, $I[n]$ penalizes errors in density variations, assigning particular weight to low-density regions where gradients are hardest to reproduce correctly. It may therefore be expected that the LPA, that is made for slowly varying densities, fails.
The error trends in the lower panel of Fig.~\ref{fig:observables} reflect this: our starting point, the LPA, has an error of over $400\%$. Already, the linear approximation, which adds information about density variations, does much better, dividing the error by more than a factor of 10. In helium, all COT approximations further reduce this error: the simplest COT1 with a percentage error of $-2.25\%$ has the smallest error, which suggests error cancellation that may be fortuitous. The COT1-av with a percentage error of $-16.28\%$ is significantly higher, but it still constitutes more than an order-of-magnitude improvement with respect to the LPA. COT1-$\alpha$ with percentage error of $-8.0\%$ performs, not surprisingly, better. The fact that it does not reduce the error even more, in spite of the parameters, is likely related to the fit, which is only optimized on a line, and not on the volume. This could, of course, be changed for an application in practice. For silicon, we find LPA overestimates $I[n]$ by 213\%, while both COT1-$\lambda$ and COT1-$\alpha$ bring the error down to $\sim$~27~\%.  Altogether, these results demonstrate that the errors made on quantities calculated with the density obtained in our orbital-free way can be understood, because our developments are systematic. They may, for certain applications, already be acceptable. In any case, they confirm that systematic improvement can be obtained, and at which places further work should lead to progress. 

The results on $I[n]$ motivate an examination of the impact on orbital-free kinetic energy functionals \cite{Mi2023}, since both $I[n]$ and kinetic energy approximations share sensitivity to $|\nabla n|^2/n$. Table~\ref{tab:kinetic_errors} reports the relative errors of orbital-free kinetic energy functionals evaluated on the approximate densities for helium, silicon and aluminum. Note that the reference values are obtained by evaluating these functionals on our reference density, \textit{not} from a self-consistent orbital-free DFT calculation. We discuss two representative functionals:
the Thomas-Fermi von Weizs\"{a}cker (TFvW)~\cite{TFvW-functional}, which incorporates gradient corrections through the von Weizs\"{a}cker term, and the Perdew-Constantin (PC07) ~\cite{PC07-functional}, which additionally depends on $|\nabla n|^2$ and $\nabla^2 n$, thus representing the highest level of gradient sensitivity. In the case of cubic helium, for TFvW, the LPA yields a large error of $196.00\%$, consistent with its failure on $I[n]$, since TFvW is directly sensitive to $|\nabla n|^2/n$. Already the linear response approximation (LRA), which adds information about density variations, reduces this to $27.26\%$. The COT approximations further reduce this systematically: COT1 reduces the error to $10.13\%$, COT1-av (or equivalently COT1-$\lambda$ with $\lambda$=1) to $-4.47\%$, and COT1-$\alpha$ to $2.35\%$. 
The PC07 functional is even more demanding: the LPA yields $234.60\%$, as PC07's dependence on the Laplacian and the squared gradient strongly amplifies any errors in the density's gradient structure. The LRA again reduces the error to $34.04\%$, and the same systematic trend holds: COT1 reduces the error to $13.98\%$, COT1-av to $-5.08\%$, and COT1-$\alpha$ to $2.81\%$. In the case of silicon, the LPA yields errors of $63.98\%$ and $58.85\%$ for TFvW and PC07, respectively. Interestingly, the LRA yields $-1.00\%$ and $1.09\%$, for TFvW and PC07, respectively. This and the analysis of $I[n]$ are consistent. On the other hand, the MADE error in Table \ref{tab:si_made} is higher for the LRA than for COT1 and COT1-$\lambda$ (see also Fig.~\ref{fig:si1}). This suggests that the quality of a density approximation should be assessed with multiple metrics, as different observables may respond differently to the same density error. In any case, the pattern is the same for the COT methods: COT1 significantly reduces the LPA error to $10.43\%$ and $9.16\%$ while COT1-$\lambda$ further improves the results to $7.05\%$ and $4.63\%$, and the best results are obtained with COT1-$\alpha$ which yields $6.54\%$ and $4.64\%$ for TFvW and PC07, respectively. We finally also report the results for aluminum, which show the same trend. Such a systematic improvement of the COT hierarchy on both functionals is nontrivial, since none of these approximations was fitted to kinetic energy observables, suggesting that the COT framework systematically captures the physically relevant features of the density gradient. 

The best results are obtained using a few parameters. While the single parameter $\lambda$ appears to depend on the class of material that is considered, a unique choice for the two parameters $A$ and $B$ that determine the function $\alpha_{\bf r}$ leads to improvement for all materials that we have examined. This very promising result should motivate further studies.

\begin{table}[t]
\caption{\label{tab:kinetic_errors}%
Relative errors (\%) of orbital-free kinetic energies from the TFvW and
PC07 functionals with respect to the values obtained with the reference density.}
\begin{ruledtabular}
\begin{tabular}{l
    S[table-format=-3.2] S[table-format=-3.2]
    S[table-format=-2.2] S[table-format=2.2]
    S[table-format= 3.2] S[table-format=2.2]}
 & \multicolumn{2}{c}{He} & \multicolumn{2}{c}{Si} & \multicolumn{2}{c}{Al}  \\
\cmidrule(lr){2-3}\cmidrule(lr){4-5}\cmidrule(lr){6-7}
Method & {TFvW} & {PC07} & {TFvW} & {PC07} & {TFvW} & {PC07} \\
\hline \addlinespace[3pt]
LPA            & 196.00 & 234.60 & 63.98 & 58.85 & 100.25 & 65.83 \\
LRA         &  27.26 &  34.04 & -1.00 &  1.09 &  11.48 &  9.91 \\
COT1           &  10.13 &  13.98 & 10.43 &  9.16 &  10.56 &  9.43 \\
COT1-$\lambda$ &  -4.47 &  -5.08 &  7.05 &  4.63 &   5.16 &  3.71 \\
COT1-$\alpha$  &   2.35 &   2.81 &  6.54 &  4.64 &   4.32 &  3.43 \\
\end{tabular}
\end{ruledtabular}
\end{table}

\section{The Kohn-Sham self-consistency loop}
\label{sec:ks}
Self-consistent Kohn-Sham calculations are time consuming, because the equations have to be solved many times. In order to explore whether our approximations may be helpful in this context, one has to study their behavior in a self-consistency cycle. For this purpose, we start from a constant density containing the correct number of electrons. The Kohn-Sham potential is then constructed using the LDA to approximate the xc functional, i.e., to calculate $v_{\rm xc}({\bf r};[n])$. Instead of solving the Kohn-Sham equation, we now use our approximations to determine the density as a functional of the resulting total KS potential, and we use this density as input for the next iteration. Using the LPA in this cycle was already studied in \cite{CANGI2013}; the iterations converge well, and the approach is very fast. We confirm that the outcome is reasonable, but the error of the LPA is of course reflected in the final result. However, we then continue the iteration procedure by using our best parameter-free approach COT1-av. The result for the density of helium in its equilibrium geometry is shown in the middle panel of Fig.~\ref{fig:made_accumulation_uncompressed}, and the resulting electron number in the upper panel of Fig.~\ref{fig:observables}. Interestingly, while the error in the low-density region increases by about 30\% with respect to the non-self-consistent density, it decreases significantly on and around the atom. 

Still, the increased error in the low density regime leads to a total electron number of $N_{\rm el}= 2.82$. We mitigate this problem during the KS cycle by partially imposing the correct electron number during the iterations: 
On one hand, the Hartree potential is, as always, calculated without the average density, since this divergent contribution is supposed to be canceled by the ionic potential. In other words, the average density is implicitly corrected in the Hartree potential through a constant shift of the density. The same does not hold for the xc potential, which has neither divergence nor cancellations. Still, we may use the electron number as an exact constraint and apply, at every iteration $(i)$, a shift $\Delta$, 
\[
v_\mathrm{xc}(\textbf{r};[n^{(i)}]) \rightarrow v_\mathrm{xc}(\textbf{r};[n^{(i)}-\Delta^{(i)}])
\]
where $\Delta^{(i)}$ is determined by the requirement that $n^{(i)}-\Delta^{(i)}$ must yield the correct number of electrons. When a straightforward shift would lead to slightly negative densities, we set the density to zero in those places and re-adjust the shift such as to conserve the electron number. Note that the shift is \textit{not} performed to adjust the final density, which therefore still has a wrong electron number, but only the density used as input to the xc potential. More details can be found in Appendix \ref{sec:compdet}.

Even when one requires a charge density with better precision than what our approximations can yield, these approximations may still be used to create an efficient starting point for standard Kohn-Sham self-consistent calculations. One simple starting point that is available today for such calculations is the density obtained by superposing the charge density of isolated (pseudo-)atoms. This is of course a very good guess for helium, but it becomes less realistic as soon as charge is moved between atoms. Silicon is a good example for such a case. Figure \ref{fig:pseudo} in Appendix \ref{app:pseudo} shows that the density obtained as the superposition of atoms is qualitatively wrong in the (110) direction. This error is remedied by COT1. A few self-consistency steps of COT1 may therefore yield a cheap and beneficial starting point for full Kohn-Sham calculations.

\section{Conclusion and outlook}
\label{sec:conclusions}

To summarize, we propose a strategy to develop increasingly precise approximations for the density as functional of a total KS potential. This strategy is based on the recently developed connector theory (COT), where an observable is given locally by the result of a model system with suitably chosen parameters. Here, our model system is the homogeneous electron gas, and we derive COT approximations by using Taylor expansions in terms of the KS potential. 

Of course, the solution of the KS equation defines by itself the density as an implicit functional of the potential. The aim of the present work is threefold: first, to design \textit{explicit} functionals. This may be useful in the context of materials design, and it can even lead, in some cases, to functionals that can be inverted and therefore open the possibility for inverse design. Second, evaluating these functionals should be fast, i.e., much faster than evaluating the exact expression. At the same time, the third requirement is, of course, sufficient accuracy: what this means is defined by the use that is made of the results.

The first requirement is automatically met in the framework of COT, and the resulting expressions are simple enough to establish direct links between the potential and the density. The computational load is perfectly negligible for the simplest approximation, the LPA, which is equivalent to the Thomas-Fermi approximation. All other approximations involve expressions based on linear response and are combined with a final LPA-like step. The linear-response part is the most costly ingredient. However, it uses the Lindhard function of the HEG, which is known analytically. The resulting integrals scale linearly with the unit cell size: the time for computation is simply proportional to the number of points where the density should be evaluated, because the static response function decays quickly, especially for low densities, with a decay range that is independent of system size.       
The results are promising: while the LPA was already explored before, and while it is known to have severe deficiencies, our increasingly sophisticated linear-response-based approximations remove a significant part of the errors. It should be stressed that our primary benchmark system is solid helium: this is an extremely inhomogeneous material, which is a worst-case test for our approximations. Indeed, we find that by compressing helium, the results improve in those regions where the density becomes more slowly varying, and less low, which is encouraging for the application to more homogeneous materials. Our results for silicon and aluminum indeed show smaller mean absolute difference errors. It should be noted that our calculations are based on norm-conserving pseudopotentials: the pseudopotential for helium is local, but other materials will require non-local potentials within standard KS-DFT (unlike the pseudopotentials used for orbital-free DFT, which are designed to be purely local \cite{Zhou2004,XU2024,Rios-Vargas2026}). The extension of the response approach to non-local potentials is straightforward in principle, but it requires the derivation and use of a Lindhard-like non-local response function for different angular momenta, which will be left for future work. 

The efficient calculation of the charge density is important because it occurs numerous times in a self-consistent KS cycle. Our approximations behave well in this context, which indicates that they might at least be used for a certain number of initial steps, i.e., as a preconditioner. They may also be of interest in the field of orbital-free DFT. 

The present work highlights some points of principle, in particular, that one may indeed be able to design functionals of the potential directly for observables. Here, we concentrate on the density as a prototypical example, but other quantities will be interesting to explore, in particular, the energy density, which will allow one to calculate the total energy and derived quantities, such as forces. Second, it shows how one can obtain understandable, systematic and controllable approximations that remain computationally efficient by making use of results that are available from a model system, here, the Lindhard function of the HEG. For practical use that requires better accuracy, the expressions can be further optimized. In particular, in the present work, we have simulated an explicit second-order expression by introducing, respectively, one or two parameters. These parameters are fitted to one system or even only to some path in one system, and we have shown that they yield improvement for the whole system and are even transferable to systems that are similar in nature. More precisely, we have shown that a single parameter $\lambda$ is system dependent, but quite transferable within a large family: solid helium at different pressures on one side, and silicon and aluminum on the other side. On top of this, the use of a function $\alpha_{\bf r}$ that depends on two parameters brings further improvement for the helium family, and very mild improvement for the semiconductors and the metal. Interestingly, the two parameters of $\alpha_{\bf r}$ seem to be quite universal, or at least, the same for all materials examined here. One may consider this result as opening a promising direction for machine learning. One could, of course, also explicitly use the second-order response function in the HEG, at the price of increasing the computational cost; however, since also higher-order response functions in the HEG have a limited decay range, the method would still scale linearly for large systems. Moreover, the introduction of exact constraints has a long and very successful history in the framework of DFT \cite{Kaplan2023}: one may also use this for the design of potential functionals. In particular, the density has the electron number as a simple constraint. We have made partial use of this idea in the present work, in particular, to improve the results of the KS self-consistency cycle in helium, as well as to improve the silicon results; more work in this direction can be done in the future. One may also try to improve the starting point of the Taylor expansions with respect to the LPA one, in particular, to overcome problems in the low-density regime, where the linear response is very short ranged and has difficulty correcting the LPA results.  

While we stress again that this work is meant as an exploration and analysis of a direction to go, the results turn out to be promising enough to suggest that the approach might be useful to get a very cheap idea of the charge density in complex systems, as well as of some related observables. With the transferability of one or two parameters that can be used to improve the results, it could also be interesting for the study of many different configurations of strained, stressed or disordered materials. Finally, it may serve as a good starting point for further refinements, for example, as starting point of a Kohn-Sham self-consistency cycle.

At the core of our strategy is the use of model results that already contain an important part of the required knowledge. This suggests, as the most interesting extension, the design of functionals of the \textit{external} potential, instead of the KS one. Indeed, the interacting static density-density response function of the HEG is available from QMC calculations \cite{Moroni1992}, and the insight obtained from the present work will transfer to this case. Such an approach would replace not only the solution of the KS equations, but it would also overcome the need for the KS potential itself. While the accuracy obtained in the present work does not justify the claim that this will be achieved in a close future, it does allow one to consider this as a possible horizon.

\acknowledgments
This work has received state funding managed by the French National Research Agency (ANR) under the France 2030 program, reference ``ANR-22-EXES-0013".

\section*{Data availability}
The code and data that support the findings of this article are openly
available~\cite{cot-repo}.

\appendix
\section{Computational Details}
\label{sec:compdet}

\subsection{General procedure}

To compare our calculations with the benchmark, we use a reference density from a DFT calculation. Since we deal with a non-interacting model, our input is the Kohn-Sham potential taken from the DFT calculation. Of course, one could in principle use any external potential to benchmark the non-interacting case.

Our test material is solid cubic helium with lattice constant $8.016$ Bohr \cite{Schuch1961,Vanzini2022}. The DFT calculations are performed using the ABINIT package \cite{abinit}, within the Local Density Approximation (LDA) with a norm-conserving, local Troullier-Martins pseudopotential \cite{Troullier1991}. 
The cutoff energy is $15$ Hartree, which produces a $30\times30\times30$ real-space grid. This grid, as well as the Kohn-Sham potential, is then used in all further calculations. All of our analyses (number of electrons, mean absolute difference error, etc.) are calculated with this grid. Densities in the figures are instead shown on a finer grid ($90\times90\times90$) produced by a $150$ Hartree plane-wave cutoff energy, for the sake of clarity of the presentation. The density is converged for both cutoff energies.

For our second material, Si, we use a lattice constant of $10.263$ Bohr, and a norm-conserving, local pseudopotential obtained from \cite{OFDFT_repo,Zhou2004} which captures the same charge density as the one obtained from a non-local pseudopotential. The cutoff energy is 12.5 Hartree, which produces a $24\times24\times24$ real-space grid. This grid, as well as the Kohn-Sham potential, is then used in all further calculations. All of our analysis is performed with this grid. For Al, the local pseudopotential is obtained from the same repository. The lattice constant is $7.652$ Bohr, the cutoff energy is 12.5 Hartree, which produces a $20\times20\times20$ real-space grid.

On the model-system side, since we work with the non-interacting HEG, both the density $n^h$ and the static linear response function $\chi_0^h$ are known analytically in real and in reciprocal space \cite{Giuliani2005}. These analytic forms are used to evaluate all model quantities throughout the calculation.

The charge density in the model is well known,

\begin{equation}\label{eqn:model}
    n^h(v^h) =\frac{k_F^3}{3\pi^2}= \frac{(-2v^h)^{3/2}}{3\pi^2}
\end{equation}
where we express the Fermi wavevector of the gas as
\begin{equation}\label{eqn:mu}
    \mu = \frac{k_F^2}{2}+v^h \implies k_F=\sqrt{2(\mu-v^h)},
\end{equation}
and set $\mu=0$ within the model system. Moreover, the static linear response function is \cite{Giuliani2005}

\begin{equation}\label{eqn:chi_analytical}
\chi^h_0(r,v_0)= -\frac{2k_F^4}{\pi^3}\frac{\sin(2k_Fr) -2k_Fr\cos (2k_Fr)}{(2k_Fr)^4}.
\end{equation}

\begin{figure}[ht]
    \centering
    \includegraphics{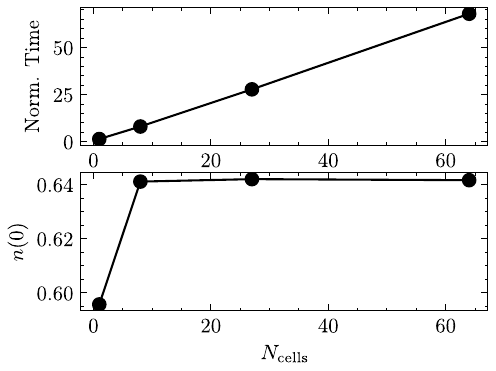}
    \caption{Linear scaling of the computational time with respect to the number of points in real space where the density is evaluated, illustrated for cubic helium. Top panel: normalized time for the calculation as a function of the number of supercells included in the calculation of the periodized response function. Bottom panel: convergence of the charge density at the atom as a function of the number of supercells.}
    \label{fig:scaling}
\end{figure}

We will use the periodized form of $\chi^h_0$ since any integral over real-space involving the response function and a periodic potential reduces to an integral over the unit cell, provided a periodized response function is used. To see this, we start from
\[
\int_{\boldsymbol{R}^3} d\textbf{r}'\, \chi_0(\textbf{r}, \textbf{r}')\, v(\textbf{r}'),
\]
we exploit the periodicity of the system by writing $\boldsymbol{R}^3 = \bigcup_{\textbf{R}} (\Omega + \textbf{R})$, where $\Omega$ is the unit cell. Changing variables to $\textbf{r}' = \textbf{r}'' + \textbf{R}$ yields
\[
\sum_{\textbf{R}} \int_{\Omega} d\textbf{r}''\, \chi_0(\textbf{r}, \textbf{r}'' + \textbf{R})\, v(\textbf{r}'').
\]
Defining the periodized response function
\[
\chi_0^{\text{per}}(\textbf{r}, \textbf{r}'') = \sum_{\textbf{R}} \chi_0(\textbf{r}, \textbf{r}'' + \textbf{R}),
\]
we obtain the final equivalency:
\[
\int_{\boldsymbol{R}^3} d\textbf{r}'\, \chi_0(\textbf{r}, \textbf{r}')\, v(\textbf{r}') = \int_{\Omega} d\textbf{r}'\, \chi_0^{\text{per}}(\textbf{r}, \textbf{r}')\, v(\textbf{r}').
\]
We have observed that the results converge with inclusion of 8 unit cells, i.e., taking into account just the nearest-neighbor cell in each direction which corresponds to cutting of the Lindhard function at $16.032$ Bohr. The finite extension of $\chi_0$ in real space implies that the method has an overall linear scaling with the number of unit cells, as shown in Fig.~\ref{fig:scaling}.

With these quantities, and using the effective potential of the real material, cubic helium, all calculations of this study are carried out straightforwardly. Throughout our work, the real space formalism is used, as it is more suitable for position-dependent expansion potentials $v_{0\mathbf{r}}$ used in the connector framework. 

All calculations presented in this work were implemented in a custom Python code. 
As long as the correct total number of electrons is not imposed, the density is calculated independently at each point ${\bf r}$, which allows one to parallelize the calculations naturally. 
The most complicated case is the COT1-av approximation, where the connector potential appears on both left- and right-hand sides of \eqref{eqn:vc-cl-con}. In the present work, we have determined the solution by iteration,

\begin{equation}
    v_{\textbf{r}}^{c,(i+1)} = \frac{1}{\chi^h_0\left(\frac{v(\textbf{r})+v_{\textbf{r}}^{c,(i)}}{2}\right)}I_\mathrm{\textbf{r}}\,,
\end{equation}
where $I_{\bf r}$ indicates the linear response calculation, which has to be performed only once, and 
 $\chi_0^h$ is to be understood as the integral in real space that appears in the denominator of \eqref{eqn:vc-cl-con}. For our case, the connector converges rather quickly, in less than 10 steps, with the initial choice $v_{\textbf{r}}^{c,(0)}=0^-$, a very small negative number.

\subsection{Corrections}
In addition to the constant shift of density mentioned in the main text, which is used as exact constraint, it may happen that the connector potential becomes locally positive, which would lead to an imaginary density and therefore also requires a correction. However, since the connector potential is essentially a suitable average of the KS potential, this problem arises only marginally, and only at a few points. We deal with it by replacing positive connector potentials by a very small negative number. 

\section{Linear response function}

\begin{figure}[h!]
    \centering
    \includegraphics{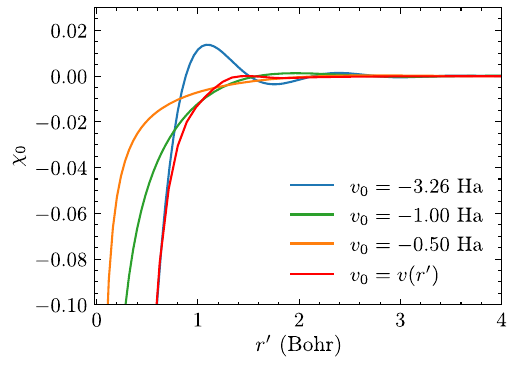}
    \caption{Lindhard function as a function of $r'=|{\bf r}-{\bf r}'|$ with ${\bf r} = 0$, for different values of the external potential, as indicated in the Legend. For the red curve, $\chi_0(r')$ has to be understood as $\chi_0(r';v_0=v({\bf r}'))$, where $v$ is the local potential for a helium atom sitting at the origin.  }
    \label{fig:response}
\end{figure}

The Lindhard function $\chi_0$ plays a central role in our approach. Figure \ref{fig:response} shows $\chi_0$ as a function of distance for different densities. The decay range is especially short for low densities, which explains why straightforward linear response is in general not sufficient. 

\section{Adjusting the chemical potential to impose exact constraints}\label{app:mu}
In extended systems, the density depends only on the difference between the local potential and the chemical potential,
\begin{equation*}
    n^h(v-\mu)=\frac{\left[2(\mu-v)\right]^{3/2}}{3\pi^2},
\end{equation*}
and the LPA and LRA densities are written as
\begin{align}
    n^{\rm LPA}(\mathbf r)
    &= n^h(v_0-\mu), \nonumber \\
    n^{\rm LRA}(\mathbf r)
    &= n^h(v_0-\mu) \nonumber \\
    &+ \int d\mathbf r'\,
    \chi_0^h(|\mathbf r-\mathbf r'|,v_0-\mu)
    \left[v(\mathbf r')-v_0\right]. \nonumber
\end{align}
The parameter $\mu$ therefore fixes the HEG reference used in the local expansion. In a periodic solid such as Si, the absolute zero of the Kohn-Sham potential is a gauge choice. Changing $\mu$ is equivalent to applying a constant shift to the local potential. This freedom can be used to remove the low-density inversion observed when the Kohn-Sham potential crosses the Hartree potential. We choose $\mu$ such that, in the crossing region, the shifted Kohn-Sham potential lies below the Hartree potential on average. In practice we use $\mu =\left\langle v_{\rm KS}(\mathbf r)-v_{\rm H}(\mathbf r)\right\rangle_{\mathbf r\in\mathbf r_{\rm cr}}$, where $\mathbf r_{\rm cr}$ denotes the region where the unshifted potentials cross.

For COT there are two HEG references. The first one, controlled by $\mu$, enters the response function used to construct the connector. The second one, controlled by $\mu_2$, enters the final HEG density:
\begin{equation*}
    n^{\rm COT}(\mathbf r)
    =
    n^h(v_{c\mathbf r}-\mu_2).
\end{equation*}
After fixing $\mu$ by the criterion above, $\mu_2$ is chosen only to impose the exact electron number,
\begin{equation*}
    \int_\Omega d\mathbf r\,n^{\rm COT}(\mathbf r)=N .
\end{equation*}
Thus $\mu$ fixes the physically useful gauge of the expansion, while $\mu_2$ enforces normalization. Such an additional freedom is absent for He. In fact, the KS potential for helium is already very close to zero, especially in the low-density region. Therefore, a further shift of the potential is forbidden, as it would lead to an imaginary density. Therefore, for He, we use $\mu=\mu_2=0$ in all calculations.

\section{COT1-$\lambda$ optimization} \label{sec:app_lambda}
\begin{figure}[h!]
    \centering
    \includegraphics[width=\columnwidth]{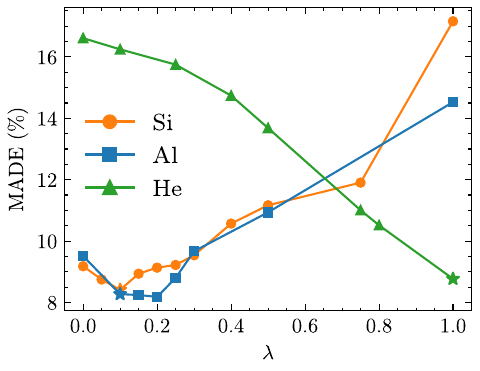}
    \caption{Mean absolute difference error (MADE) as a function of the parameter $\lambda$ for the COT1-$\lambda$ approximation for the materials considered in this work. The $\lambda$ values chosen for the final results are indicated by stars. The green curve is divided by $4$ for better visibility.}
    \label{fig:lambda_scan}
\end{figure}

We present here the $\lambda$ optimization procedure for the materials considered in the main text. The MADE obtained (through the unit cell) for a series of values for $\lambda$ is shown in Fig.~\ref{fig:lambda_scan}. We see that the optimal $\lambda$ is around the same value for Si and Al, whereas the opposite is true for He. The $\lambda$ values chosen for the final results are indicated by stars in the figure.

\section{COT1 versus charge density from superposition of atomic densities}\label{app:pseudo}

\begin{figure}[h!]
    \centering
    \includegraphics{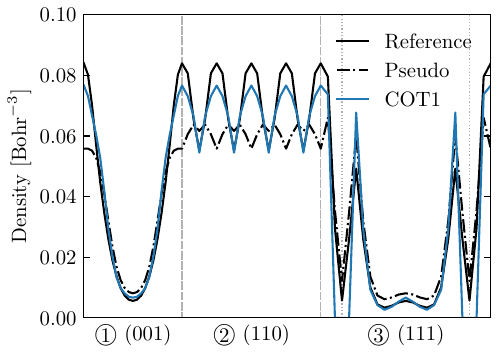}
    \caption{Charge density along the path in Si obtained from superposing atomic densities (dash-dotted black), versus the COT1 approximation (blue). The solid black is the benchmark density.}
    \label{fig:pseudo}
\end{figure}

In this section, we compare the COT1 approximation with the charge density obtained from a superposition of atomic densities obtained from the pseudopotential. Figure \ref{fig:pseudo} compares the benchmark density along the path and the simple and parameter-free COT1 density with the density obtained by superposing atoms. We also show the LPA for reference.

\bibliography{main}

\end{document}